\documentclass[lettersize,journal]{IEEEtran}
\usepackage{amsmath,amsfonts}
\usepackage{algorithm2e}
\usepackage{array}
\usepackage[caption=false,font=normalsize,labelfont=sf,textfont=sf]{subfig}
\usepackage{textcomp}
\usepackage{subcaption}
\usepackage{stfloats}
\usepackage{url}
\usepackage{verbatim}
\usepackage{graphicx}
\usepackage{cite}
\usepackage{graphicx} % Add this package in the preamble
\usepackage{amsmath}
\usepackage{amsmath}
\usepackage{amsmath, amssymb}
\usepackage{xcolor}	
\usepackage{amsmath} 
\usepackage{makecell}
\usepackage{algpseudocode}
\usepackage{graphics}
\usepackage{makecell}
\usepackage{multirow}
\usepackage{amsmath, amsthm}

\usepackage{amssymb}
\usepackage{amsmath}
\usepackage{subfig}  % Include this in your document preamble
\usepackage{enumitem}
\newcounter{assumption} % assuming you want continuous numbering throughout the document
\usepackage{yfonts}

\newcounter{remark}

\newcounter{theorm} % Continuous numbering throughout the document

\newcounter{lemma}

\newcounter{Corollary} % Assuming you want continuous numbering throughout the document

\begin{document}
	
	\title{Robust Resource Allocation in RIS-Assisted Wireless Networks Integrating NOMA and Over-the-Air Federated Learning

    %RIS-Aided Hybrid Wireless Networks Integrating NOMA and Over-the-Air Federated Learning%: A Deep Reinforcement Learning Approach
    }
	
	%\author{IEEE Publication Technology,~\IEEEmembership{Staff,~IEEE,}
		%}
	
	% The paper headers
	%\markboth{IEEE Transactions on Vehicular Technology, January~2024}%

	\author{ Saeid Pakravan, Mohsen Ahmadzadeh, Ming Zeng, Ghosheh Abed Hodtani, Xingwang Li, Ji Wang, and Gongpu Wang
		
		\thanks{S. Pakravan and M. Zeng are with the Department of Electric and Computer Engineering, Laval University, Quebec City, QC, CA. email: saeid.pakravan.1@ulaval.ca;  ming.zeng@gel.ulaval.ca.}

\thanks{M. Ahmadzadeh and G. Abed Hodtani are with the Department of Electric and Computer Engineering, Ferdowsi University, Mashhad, Iran. email: m.ahmadzadehbolghan@mail.um.ac.ir; hodtani@um.ac.ir.}

\thanks{X. Li is with the School of Physics and Electronic Information
Engineering, Henan Polytechnic University, Jiaozuo 454000, China. email:
lixingwang@hpu.edu.cn.}

\thanks{Ji Wang is with the Department of Electronics and Information Engineering, College of Physical Science and Technology, Central China Normal University, Wuhan 430079, China. email: jiwang@ccnu.edu.cn.}

\thanks{Gongpu Wang is with the Beijing Key Laboratory of Transportation Data
Analysis and Mining, Beijing Jiaotong University, Beijing, China. email: gpwang@bjtu.edu.cn.}

}

	\maketitle

	\begin{abstract} 

This paper addresses the critical issue of spectrum scarcity and the need to support diverse services, including communication and learning tasks, by presenting a reconfigurable intelligent surface (RIS)-aided wireless network framework that integrates non-orthogonal multiple access (NOMA) with over-the-air federated learning (AirFL). The proposed system leverages RIS’s ability to adaptively shape wireless channels, aiming to enhance overall network performance for both communication and learning through concurrent uplink transmissions. To tackle critical challenges, such as co-channel interference, imperfect channel state information (CSI) and successive interference cancellation (SIC), we develop an optimization framework that focuses on minimizing the optimality gap. This joint optimization is formulated as a non-convex problem, complicated by the intricate interactions between NOMA and AirFL users and the impact of imperfect CSI and SIC. To overcome these challenges and reduce the optimality gap, we reformulate the optimization as a Markov decision process and solve it using a long short-term memory deep deterministic policy gradient (LSTM-DDPG) algorithm, a memory-based approach within the realm of deep reinforcement learning (DRL). Simulation results demonstrate that the proposed approach achieves faster convergence, lower variance, and improved robustness under channel uncertainty, outperforming baseline DRL algorithms such as DDPG, soft actor-critic (SAC), and advantage actor-critic (A2C).

	\end{abstract}
	

	\section{Introduction}
	
	The exponential growth of internet of things (IoT) devices, coupled with advancements in artificial intelligence (AI), is driving a significant transition from the ``connected everything'' paradigm of the fifth generation (5G) to the ``connected intelligence'' of the beyond 5G (B5G) and sixth generation (6G) systems\cite{letaief2019roadmap}. This evolution extends beyond simple data transmission to encompass a wide range of services, including sensing, communication, and edge intelligence, critical for applications like augmented reality, connected vehicles, advanced home automation, and intelligent IoT devices \cite{saad2019vision}. However, this shift also introduces substantial challenges, particularly in ensuring data privacy, managing limited communication resources, and efficiently handling distributed big data \cite{hu2022ris}.
	
	Federated learning (FL) has gained attention as an effective solution to mitigate these challenges by preserving data privacy and minimizing communication overhead through decentralized model training across edge devices \cite{sery2021overtheair}. It ensures that sensitive data remains localized on each device, while only model updates are shared and aggregated centrally, thereby safeguarding privacy and significantly reducing the need for large-scale data transmission. Building on this approach, recent advancements have introduced over-the-air federated learning (AirFL), an innovative method that leverages the superposition properties of wireless multiple access (MAC) channels to enhance the efficiency of model aggregation. In AirFL, the model updates from edge devices are aggregated through over-the-air computation (AirComp). This technique capitalizes on the interference properties inherent in wireless channels, allowing the direct combination of analog signals. By doing so, AirFL obviates the need for individual parameter decoding, thereby reducing latency and improving spectrum efficiency within shared time-frequency resources. This approach not only streamlines the aggregation process but also optimizes the use of available communication resources, contributing to a more efficient and scalable FL framework \cite{8952884, pakravan2024robust, 11111711}.
	
	As spectrum resources become increasingly constrained and the demand for higher bandwidth intensifies, integrating distributed machine learning with advanced wireless communication technologies is crucial for optimizing network performance and ensuring seamless operations in B5G environments \cite{zhang20196g, 10185552}. In this context, non-orthogonal multiple access (NOMA) has gained prominence as a technique to enhance spectral efficiency. By allowing multiple users to share the same frequency band through power-domain differentiation, NOMA significantly increases capacity and facilitates efficient resource allocation, particularly in scenarios with diverse user demands \cite{8972353}. The integration of NOMA with FL frameworks is critical for addressing the growing demands of next-generation wireless networks.
	
	Another significant advancement in this domain is the emergence of reconfigurable intelligent surfaces (RIS), which has garnered substantial attention for their potential to enhance wireless network performance \cite{9360709, 10363658, 10923651,  8741198, mmmnew}. RIS leverages a large array of low-cost passive elements to dynamically adjust the wireless environment by tuning signal phase shifts. This capability effectively mitigates channel fading and interference, thereby improving signal reception. By combining reflected signals with direct paths, RIS increases the received signal power and reduces the required transmit power. Furthermore, RIS enables advanced passive beamforming techniques to manage interference more effectively. These enhancements contribute significantly to the optimization of next-generation wireless networks, supporting higher data rates, improved coverage, and overall better performance.
	
	The symbiotic relationship between wireless networks, NOMA, AirFL, and RIS not only alleviates resource limitations but also fosters enhanced communication efficiency, preservation of data privacy, and enhanced model training across distributed edge devices. This integration signifies a pivotal advancement toward the seamless deployment of FL in wireless environments, promising increased scalability, resilience, and adaptability within resource-constrained settings. Such a comprehensive framework sets the stage for the robust implementation of FL methodologies in wireless applications, ushering in a new era of intelligent connectivity. While combining multi-RIS, NOMA, and AirFL introduces additional complexity, it is mitigated through the proposed optimization method, which avoids prohibitive computational overhead. The integration is especially advantageous in dense and resource-constrained networks, where improved spectral efficiency, coverage, and low-latency learning are essential.

	In this paper, we tackle the critical issue of optimizing the performance of RIS-aided wireless networks, specifically focusing on the integration of NOMA with AirFL. Our proposed framework harnesses the capabilities of deep reinforcement learning (DRL) \cite{ sheikhzadeh2021ai} to jointly manage beamforming vector, power allocation, and RIS configuration in the face of channel uncertainties and heterogeneous resource constraints. Our approach aims to enhance communication quality and learning performance in hybrid networks, ensuring efficient spectrum utilization and robust network operation.

	\subsection{ Previous Works} 
	
	The integration of RIS, NOMA, and AirFL in wireless networks has garnered significant attention in recent years due to their potential to enhance spectral efficiency, reduce latency, and improve the overall performance of communication systems \cite{9546764, 66666, 9829190, 111222, 444444, 10227329, 9798757, 9815289, 10379484, 10093073, 10283543}. Various studies have explored these technologies both individually and in combination to address the challenges posed by next-generation wireless networks.
	
	In \cite{9546764}, the authors investigated the use of RIS to enhance AirComp systems by dynamically reconfiguring the wireless channel environment. Their findings showed that RIS effectively mitigates channel fading, thus enhancing the accuracy of over-the-air (OTA) aggregation and boosting FL performance. \cite{66666} studied the deployment and passive beamforming design in RIS-enhanced massive NOMA networks, demonstrating that RIS optimizes signal reception and manages interference among users with different power levels, ultimately improving spectral efficiency. Another study in \cite{9829190} examined the trade-offs between accuracy and integrity in RIS-aided AirFL systems, proposing an optimization framework that balances model accuracy with system integrity, thereby underscoring RIS's role in maintaining network stability under diverse conditions.
	
	Further extending these investigations, \cite{111222} focused on joint location and beamforming design in simultaneously transmitting and reflecting RIS (STAR-RIS)-aided NOMA systems, highlighting RIS's ability to optimize both coverage and beamforming in complex network environments. Additionally, \cite{444444} analyzed the impact of large-scale STAR-RIS deployment in randomly distributed wireless networks, demonstrating its potential to enhance system performance by leveraging its inherent unpredictability for improved signal manipulation and network adaptability. \cite{10227329} introduced a DRL-based approach to optimize energy efficiency (EE) in simultaneous wireless information and power transfer (SWIPT)-enabled AirFL systems. Their results indicated that RIS integration could minimize energy consumption in edge devices while preserving learning accuracy, showcasing the potential of RIS for energy-constrained IoT networks. In \cite{9798757}, researchers explored the integration of AirFL and NOMA within RIS-assisted networks, proposing a framework that leverages RIS for enhanced spectrum utilization and learning efficiency, demonstrating its critical role in managing co-channel interference and optimizing resource allocation. Expanding on this, \cite{9815289} proposed a comprehensive framework for STAR-RIS-integrated NOMA and AirFL systems, providing insights into the optimization and performance analysis of such complex networks.

	Recent studies have also examined the integration of wireless power transfer with AirFL. In \cite{10379484}, the authors introduced strategies that combine RIS and SWIPT to support efficient OTA model aggregation in FL, reflecting a growing interest in power-efficient designs for next-generation networks. Additionally, \cite{10093073} investigated the deployment of multiple RIS in NOMA-assisted FL, focusing on latency reduction and auction-based resource allocation, which advances practical deployment in large-scale networks. Finally, the study in \cite{10283543} tackled the challenges of noisy downlink communication in RIS-assisted OTA adaptive FL, presenting adaptive strategies to mitigate the detrimental effects of noise. This study further highlights RIS's importance in sustaining reliable communication under adverse conditions.

	Despite these advances, existing works often focus on optimizing one or two aspects of the system either RIS configuration, NOMA efficiency, or AirFL performance, without fully exploring the potential of a holistic approach that integrates all three technologies. Moreover, conventional model-based optimization approaches, such as successive convex approximation and alternating optimization, have been widely used in related problems. However, their applicability is limited in this context due to the high-dimensional, non-convex, and dynamic nature of the joint optimization problem, motivating the use of learning-based solutions.

    To further clarify the novelty and positioning of the proposed framework, a comparison with representative existing works is summarized in Table~I.

\begin{table*}[!t]
\caption{Summary and Comparison of Related Works} \label{tab:table6}
\centering

\begin{tabular}{|p{0.42cm}|p{1.7cm}|c|c|c|c|c|c|p{5.3cm}|p{1.5cm}|}
\hline
Ref & Performance Metrics & \multicolumn{6}{c|}{System Model} & Main Contribution & Optimization Method \\
\cline{3-8}
&  & NOMA & OTA & FL & RIS & STAR-RIS & Uncertainty & & \\
\hline
\cite{9546764} &  Minimize MSE &  $\times$ & \checkmark & $\times$ & \checkmark & $\times$ & $\times$ & Proposed an RIS-assisted AirComp system to overcome worst channel conditions. & SCA and MPM\\
\hline
\cite{66666} &  Maximize EE &  \checkmark & $\times$ & $\times$ & \checkmark & $\times$ & $\times$ & Developed a RIS-NOMA framework with machine learning for tele-traffic prediction. & DRL\\
\hline
\cite{9829190} &  Minimize MSE & $\times$ & \checkmark & \checkmark & \checkmark & $\times$ & \checkmark& Proposes an AirFL framework that ensures accuracy and integrity via optimized beamforming and RIS phase shifts. & SCA and DC \\
\hline
\cite{111222} & Maximize weighted sum rate & \checkmark & $\times$ & $\times$ & $\times$ & \checkmark & $\times$& Analyzes the optimal coordination of beamforming and location in STAR-RIS-assisted downlink networks. & SCA and SDP\\
\hline
\cite{444444}   & Average detection error probability & $\times$ & $\times$ & $\times$ & $\times$ & \checkmark & \checkmark & Investigates the impact of large-scale STAR-RIS deployment in randomly distributed wireless networks. & SDP\\
\hline
\cite{10227329} &  Maximize EE &  $\times$ & \checkmark & \checkmark & $\times$ & $\times$ & $\times$ & Develops a SWIPT-based AirFL framework for EE distributed learning in IoT networks. & DRL\\
\hline
\cite{9798757} &  Maximize hybrid rate &  \checkmark & \checkmark & \checkmark & \checkmark & $\times$ & $\times$ & Proposes an RIS-assisted hybrid network integrating AirFL and NOMA for optimized hybrid rates. & DC, SCA, and SDP\\
\hline
\cite{9815289} & Minimize optimality gap & \checkmark & \checkmark & \checkmark & $\times$ & \checkmark & $\times$ & Proposes a novel approach to tackle spectrum scarcity, optimize uplink communication, and assess AirFL convergence.& SCA and SDP\\
\hline
\cite{10379484} & Minimize error gap & $\times$ & \checkmark & \checkmark & \checkmark & $\times$ & $\times$ & Proposes a RIS-assisted SWIPT-AirFL system, optimizing device selection and beamforming for improved convergence and accuracy. & SCA\\
\hline
\cite{10093073} & Minimize latency& $\times$ & \checkmark & \checkmark & \checkmark & $\times$ &$\times$ & Presents a model where an RIS provider deploys multiple RISs to enhance FL in nearby BSs using uplink NOMA. & SCA and SDP\\
\hline
\cite{10283543} & Optimize computation resources & $\times$ & \checkmark & \checkmark & \checkmark & $\times$ & \checkmark& Proposes an adaptive RIS-assisted FL approach that outperforms existing designs under imperfect CSI and heterogeneous data.& SCA\\
\hline
Our work & Minimize optimality gap & \checkmark & \checkmark & \checkmark & \checkmark & $\times$ & \checkmark& Proposes an RIS-aided hybrid network integrating AirFL and NOMA under imperfect CSI and SIC. & DRL \\
\hline
\end{tabular}

\vspace{5pt}
\text{MSE:} Mean square error,
\text{SCA:} Successive convex approximation, 
\text{MPM:} Mirri-Prox method,
\text{DC:} Difference of convex functions programming, 
\text{SDP:} Semidefinite programming,
\text{$\checkmark$:} Item is supported, and \text{$\times$:} Item is not supported.

\end{table*}

	\subsection{Motivations and Contributions}
	
	The motivation behind this research is to develop a comprehensive framework that integrates RIS, NOMA, and AirFL, thereby leveraging their complementary strengths to enhance communication efficiency and learning performance in next-generation wireless networks. At the same time, it is essential to carefully manage the resulting system complexity, which motivates the use of learning-based optimization approaches in this work. This framework offers a more robust and efficient solution to the key challenges faced by future wireless systems, addressing limitations inherent to each individual technology.

However, the performance of RIS-assisted systems is fundamentally influenced by imperfections in channel state information (CSI) and successive interference cancellation (SIC), which can significantly degrade system efficiency. Imperfect CSI, resulting from estimation errors and feedback delays, leads to suboptimal beamforming and increased interference, while imperfect SIC affects the accurate decoding of overlapping signals in non-orthogonal transmissions. These challenges become even more pronounced in the presence of hybrid users, which simultaneously engage in multiple communication tasks, further complicating resource allocation and interference management. Thus, addressing the limitations imposed by imperfect CSI and SIC is crucial for ensuring reliable and efficient operation in RIS-assisted networks.

To this end, this paper analyzes the impact of imperfect CSI and SIC in RIS-assisted systems with hybrid users and proposes a joint optimization framework to mitigate their effects. We derive the optimality gap for a simplified system model to quantify performance degradation and develop strategies to minimize this gap. By analyzing these impairments, we provide insights into their impact on network performance and propose solutions to enhance system reliability.
	
	To address the optimization problem associated with this issue, we employ DRL, a state-of-the-art approach known for its effectiveness in dynamic and uncertain environments. DRL leverages neural networks to learn optimal policies through interaction with the environment, making it an ideal tool for real-time optimization in complex systems \cite{10227329, 10109153, saeid111, 11106811, 10622931}. By applying DRL, we aim to adaptively and efficiently optimize the RIS-aided system, ensuring that quality of service (QoS) requirements are met despite the presence of imperfect CSI and SIC.

	The key contributions of this paper can be outlined as follows:
    %The key contributions of this paper are as follows:
	
	\begin{itemize}
		\item  
        
        \textit{Framework development and convergence analysis:} We propose an RIS-aided network as a universal framework that simultaneously supports both communication-focused users and learning-oriented AirFL users through a single concurrent transmission. Learning-oriented AirFL users are devices engaged in AirFL, where instead of conventional data transmission, they participate in a distributed learning process by transmitting local model updates. These updates are aggregated directly over the wireless channel using the principles of analog superposition, enabling efficient and low-latency global model training. To assess the performance of this framework, we conduct a comprehensive convergence analysis. Specifically, we investigate the optimality gap between the actual loss and the optimal loss, deriving a closed-form expression that quantifies the impact of aggregation errors, particularly the MSE of model aggregation, on the convergence behavior of the AirFL algorithm.
		
		\item \textit{Optimality gap minimization and DRL solutions:} We formulate an optimization problem that explicitly incorporates the impact of imperfect CSI and SIC, establishing a framework to address these challenges effectively. Given the complexity of this problem, we reformulate it as a Markov decision process (MDP) to facilitate the use of dynamic optimization techniques. To solve the MDP, we employ DRL methods, which are well-suited for environments with time-varying conditions. Specifically, we introduce the long short-term memory deep deterministic policy gradient (LSTM-DDPG) algorithm,which captures temporal correlations and enables real-time decision-making to enhance system adaptability and performance.
		
	\item \textit{Performance evaluation:} To evaluate the effectiveness of the proposed AI-enhanced RIS-assisted hybrid network, we perform extensive simulations using both synthetic and real-world datasets. The results demonstrate that this framework effectively manages concurrent uplink communication and FL tasks. The LSTM-DDPG algorithm consistently outperforms other benchmarks, including DDPG, soft actor-critic (SAC), and advantage actor-critic (A2C). Furthermore, the integration of RIS significantly improves the robustness and efficiency of the proposed network design.

	\end{itemize}

	\subsection{Notations}
	
    Italic and boldface letters denote scalars and vectors, respectively. The transpose operation is indicated by $(\cdot)^T$, while $(\cdot)^H$ signifies the conjugate transpose. The operator $\mathbb{E}[\cdot]$ denotes expectation. $\Re$ denotes the real part of a complex number, $|\cdot|$ indicates the cardinality of a set or the absolute value of a scalar, and $\|\cdot\|$ refers the Euclidean norm. The symbol $\hat{\cdot}$ signifies an estimate, and $\text{diag}(\cdot)$ represents a diagonal phase-shift matrix.

	\section{System Model}

	\subsection{Network and Channel Model}

	As depicted in Fig. 1, we investigate a wireless network enhanced by \( X \) RISs, each comprising \( M_x \) reflecting elements, where \( x \in \{1, 2, \ldots, X\} \). The network consists of a single-antenna base station (BS) serving hybrid users, which are classified into NOMA and AirFL users. Both user types employ non-orthogonal communication schemes over a shared wireless MAC. The hybrid users are represented by the set \(\mathcal{I} = \{1, 2, \ldots, K, K + 1, K + 2, \ldots, K + N\}\), where AirFL users are indexed by \(\mathcal{K} = \{1, 2, \ldots, K\}\) and NOMA users by \(\mathcal{N} = \{K + 1, K + 2, \ldots, K + N\}\). For the \( x \)-th RIS, the reflection matrix is defined as 
	$\Theta_x = \text{diag}(e^{j\theta_{x,1}}, e^{j\theta_{x,2}}, \ldots, e^{j\theta_{x,M_x}})$,
	where \(\theta_{x,m}\) indicates the phase shift introduced by the \( m \)-th element of the \( x \)-th RIS.\footnote{The RIS is modeled using an ideal diagonal phase-shift matrix with independently controllable elements. In practice, mutual coupling between adjacent elements may introduce interdependence and degrade beamforming gain. These effects are not considered in this work, and thus the results represent an upper-bound performance under ideal RIS assumptions.}

	\begin{figure}[]
		\centering
		\includegraphics [width=8.7cm, height=5.5cm]{Fig1-new.png}%
		\qquad
		\caption{Illustration of the multi-RIS-assisted system integrating NOMA and AirFL, where NOMA users transmit data signals, while AirFL users upload local gradients for over-the-air aggregation.}%
		\label{fig:example}%
	\end{figure}
	
	In the proposed framework, the characterization of small-scale fading channels is outlined as follows: the channel from the 
	$x$-th RIS to the BS is represented by $\boldsymbol{G}_x \in \mathbb{C}^{M_x\times1}$, the channel from the $i$-th user to the 
	$x$-th RIS is represented by $\boldsymbol{h}_{i,x} \in \mathbb{C}^{M_x\times1}$, and the channel from the $i$-th user to the BS is indicated by $r_i \in \mathbb{C}^{1\times1}$, $\forall i \in I$. Additionally, the large-scale path loss $L(d)$ is modeled as $\varrho_0(d)^{-\alpha}$, where $\varrho_0$ represents the path loss at a reference distance of one meter, $d$ denotes the individual link distance, and $\alpha$ represents the path-loss exponent.

Following the approach in \cite{9353406, 66666}, all user-related links are modeled as Rayleigh fading channels due to the presence of rich scattering. In contrast, the BS–RIS link is assumed to experience Rician fading, attributed to the elevated positions of both the BS and the RIS.
Consequently, the channel coefficients for the link between the $x$-th RIS and the BS are given by:
	\begin{equation}
		\boldsymbol{G}_x = \sqrt{L(d_{0,x})} \left( \sqrt{\frac{\kappa}{\kappa+1}}\boldsymbol{G}_x^{\text{LoS}}+\sqrt{\frac{1}{\kappa+1}}\boldsymbol{G}_x^{\text{NLoS}} \right),
	\end{equation}
	where \(d_{0,x}\) represents the distance between the BS and the 
	$x$-th RIS, \(\kappa\) denotes the Rician factor, and $\boldsymbol{G}_x^{\text{LoS}}$ and $\boldsymbol{G}_x^{\text{NLoS}}$ represent the deterministic Line-of-Sight (LoS) and Rayleigh fading components, respectively. Specifically, the LoS component $\boldsymbol{G}_x^{\text{LoS}}$ is given by $\boldsymbol{G}_x^{\text{LoS}} = [1, e^{j(2\pi \bar{d}/\bar{\lambda}) \sin \vartheta_x^{{\text{AoD}}}}, \ldots, e^{j(2\pi \bar{d}/\bar{\lambda})(M_x-1) \sin \vartheta_x^{{\text{AoD}}}}]^{T}$, where \(\bar{d}\) denotes the element separation distance, \(\bar{\lambda}\) represents the wavelength, and \(\vartheta_x^{{\text{AoD}}}\) denotes the angle of departure, assumed to be uniformly distributed in \([0, 2\pi]\). For simplicity, $\bar{d}/\bar{\lambda}=1/2$ \cite{9090356}.
	Consequently, the composite channel from the $i$-th user to the BS, including direct and RIS-aided paths, is given by:
	\begin{equation}
		\bar{{h}}_i ={r}_i+\sum_{x=1}^{X}\boldsymbol{G}_x^H \Theta_x \boldsymbol{h}_{i,x}={r}_i+ \sum_{x=1}^{X}\boldsymbol{v}_x^H \boldsymbol{\Phi}_{i,x}, \quad \forall i \in I,
	\end{equation}
	where $\boldsymbol{\Phi}_{i,x} = \text{diag}(\boldsymbol{G}_x^H) \boldsymbol{h}_{i,x}$ and $\boldsymbol{v}_x = [v_{x,1}, v_{x,2}, \ldots, v_{x,M}]^H$ with $v_{x,m} = e^{j\theta_{x,m}}$. It is noteworthy that the vector $v_x$ effectively encapsulates the RIS phase shifts, segregating them from the reflective channels $\boldsymbol{\Phi}_{i,x}$. \footnote{The composite channel $\bar{h}_k$ implicitly includes delay-induced phase rotations caused by propagation delays. Therefore, both amplitude fading and phase misalignment effects are embedded in the effective channel representation.}

	\subsection{Federated Learning Training Process}
	
	This subsection provides a detailed overview of the training procedure in FL, outlining the sequential steps involved in each training round or communication iteration denoted by $\mathcal{T} = \{1, 2, \ldots, T\}$.
	
	\begin{enumerate}
		\item \textbf{Global Model Broadcast:} At the beginning of each training round, the BS initiates the process by broadcasting the latest version of the global model, denoted as $\boldsymbol{w}^{(t)}$. This model is shared with all AirFL users, where $\boldsymbol{w}^{(t)} \in \mathbb{R}^Q$, and $Q$ indicates the dimensionality of the parameter space of the global model. This step ensures that all AirFl users start the training round with the same global model, providing a common reference point for subsequent updates.

		\item \textbf{Local Model Update:} Subsequently, each AirFL user independently updates its local gradient $\boldsymbol{g}_k^{(t)}$ based on its individual dataset. Specifically, the local gradient of the $k$-th user is obtained as:
		\begin{equation}
			\boldsymbol{g}_k^{(t)} = \nabla F_k(\boldsymbol{w}^{(t)}) = \frac{1}{|\mathcal{D}_k|} \sum_{i=1}^{|\mathcal{D}_k|} \nabla f(\boldsymbol{w}_{k}^{(t)}, x_{k,i},y_{k,i}), 
		\end{equation}
		where $\nabla f(\boldsymbol{w}_{k}^{(t)}, x_{k,i},y_{k,i})$ is the gradient of the local loss function for sample $(x_{k,i},y_{k,i}), i \in \{1, . . . ,|\mathcal{D}_k|\}$ at user $k$. It is assumed that all local datasets have the same size, denoted as $|\mathcal{D}_k| = D$, as discussed in \cite{9292468}.

		\item \textbf{Model Aggregation:} 
		Following the computation of local gradients, all AirFL users transmit their respective gradients $\{\boldsymbol{g}_k^{(t)}\}$ to the BS via wireless channels for global synchronization. Upon receiving the local gradients from all AirFL users, the BS updates the global model using the following equation:
		\begin{equation}
			\boldsymbol{w}^{(t+1)} = \boldsymbol{w}^{(t)} - \lambda \boldsymbol{g}^{(t)}, \quad \text{with} \quad \boldsymbol{g}^{(t)} = \frac{1}{K} \sum_{k \in \mathcal{K}} \boldsymbol{g}_k^{(t)},
		\end{equation}
		where $\boldsymbol{g}^{(t)}$ represents the global gradient computed as the average of all local gradients. Here, $\lambda$ denotes the learning rate chosen by the BS.
	\end{enumerate}

	The training process continues iteratively, with each round repeating these steps until either the convergence criteria are satisfied or the predefined maximum number of iterations, denoted as $T$, is reached.

\subsection{RIS-Aided Concurrent Transmission}
	
In the $t$-th training round of RIS-assisted communication systems, simultaneous uplink transmission is facilitated by encoding local data $\boldsymbol{d}_n^{(t)}$ from the $n$-th NOMA user and the local gradients $\boldsymbol{g}_k^{(t)}$ from the $k$-th AirFL user are each encoded into communication symbols $s_n^{(t)}$ and computation symbols $s_k^{(t)}$, respectively, satisfying $\mathbb{E}\big[s_i^{(t)}\big]=0$, $\mathbb{E}\big[|s_i^{(t)}|^2\big]=1$. This encoding process ensures that the distinct types of information are properly formatted for transmission. Both NOMA and AirFL users transmit their encoded symbols concurrently over the same time-frequency resource, leveraging the assistance of RISs in addition to the direct transmission link. Consequently, the received superposition signal at the BS is expressed as:
	\begin{equation}
		{y}^{(t)} = \underbrace{\sum_{k=1}^{K} \bar{{h}}_k^{(t)} \sqrt{p_k^{(t)}} s_k^{(t)}}_{\text{AirFL users}} + \underbrace{\sum_{n=K+1}^{K+N} \bar{{h}}_n^{(t)} \sqrt{p_n^{(t)}} s_n^{(t)}}_{\text{NOMA users}} + \underbrace{{z}_0^{(t)}}_{\text{Noise}},
	\end{equation}
	where $p_k^{(t)}$ ($p_n^{(t)}$) denotes the power scaling factor at the $k$-th ($n$-th) user, satisfying $p_k^{(t)} \in\left[ 0,P_{\mathrm{max}} \right]$, with $P_{\mathrm{max}}$ being the maximum power constraint,  ${z}_0^{(t)} \sim \mathcal{CN}(0, \sigma^2)$ represents additive white Gaussian noise (AWGN) with zero mean and noise power $\sigma^2$.

	Although RISs offer remarkable advantages by facilitating simultaneous communication, they also bring about notable co-channel interference, especially in environments characterized by extensive connectivity. It is essential to recognize that the influence of this interference differs between users employing NOMA and those utilizing AirFL. For NOMA users, the presence of co-channel interference poses challenges during the decoding process, potentially leading to errors or degraded signal quality. On the other hand, AirFL users may experience a contrasting effect, as the interference can actually enhance aggregation performance by providing additional signal components. Therefore, implementing effective signal processing strategies is essential to manage co-channel interference in heterogeneous fusion networks, aiming to enhance both communication throughput and learning efficiency.

	\subsection{Imperfect CSI and SIC}

In this study, we consider the practical scenario with imperfect CSI and SIC modeled as the following.

	{\bf{Imperfect CSI:}} Within our theoretical framework, we incorporate the presence of imperfections within both the direct communication links between users and BS, as well as the cascaded communication paths encompassing users, RISs, and BS. These imperfections are represented as follows:
	\begin{equation}
{r}_i=\hat{{r}}_i+\boldsymbol{\varepsilon}_{r,i},
	\end{equation}
	\begin{equation}
\boldsymbol{\Phi}_{i,x}=\hat{\boldsymbol{\Phi}}_{i,x}+\boldsymbol{\varepsilon}_{\Phi,i,x},
	\end{equation}
	where $\hat{{r}}_i$ is the estimated direct channel, $\boldsymbol{\varepsilon}_{r,i}$ is the estimation error for the direct link, $\hat{\boldsymbol{\Phi}}_{i,x}$ denote the estimated channel from $i$-th user through the $x$-th RIS, and $\boldsymbol{\varepsilon}_{\Phi,i,x}$ represents the estimation error at the $x$-th RIS. These errors follow a circularly symmetric complex Gaussian (CSCG) distribution, i.e., $\boldsymbol{\varepsilon}_{r,i} \sim C\mathcal{N}(0,\sigma^{2}_{r,i})$ and $\boldsymbol{\varepsilon}_{\Phi,i,x} \sim C\mathcal{N}(0,\sigma^{2}_{\Phi,i,x}\boldsymbol{I})$. Here, $\sigma^{2}_{r,i}$ and $\sigma^{2}_{\Phi,i,x}$ denote the variances of the direct channel estimation and cascaded channel estimation, respectively. Therefore, considering estimation imperfections, the composite channel from the $i$-th user to the BS, $\bar{{h}}_i$, can be formulated as:
	\begin{multline}
\bar{{h}}_i=\hat{{r}}_i+\boldsymbol{\varepsilon}_{r,i}+\sum_{x=1}^{X}\boldsymbol{v}_x^H(\hat{\boldsymbol{\Phi}}_{i,x}+\boldsymbol{\varepsilon}_{\Phi,i,x})=\hat{{h}}_i+\boldsymbol{\varepsilon}_{h,i},
	\end{multline}
	where $\hat{{h}}_i = \hat{{r}}_i +\sum_{x=1}^{X} \boldsymbol{v}_x^H\hat{\boldsymbol{\Phi}}_{i,x}$, $\boldsymbol{\varepsilon}_{h,i} = \boldsymbol{\varepsilon}_{r,i} + \sum_{x=1}^{X}\boldsymbol{v}_x^H\boldsymbol{\varepsilon}_{\Phi,i,x}$, and $\bar{{h}}_i\sim C\mathcal{N}(\hat{{h}}_i,\sigma^{2}_{h,i}), \ \forall i \in I$. Due to the lack of correlation among estimation errors for different channels, the variance of the composite channel $\sigma^{2}_{h,i}$ is determined by:
	\begin{multline}
		\sigma^{2}_{h,i} = \mathbb{E}((\boldsymbol{\varepsilon}_{r,i} + \sum_{x=1}^{X}\boldsymbol{v}_x^H\boldsymbol{\varepsilon}_{\Phi,i,x})(\boldsymbol{\varepsilon}_{r,i} + \sum_{x=1}^{X}\boldsymbol{v}_x^H\boldsymbol{\varepsilon}_{\Phi,i,x})^{H}) 
		\\= \mathbb{E}(\boldsymbol{\varepsilon}_{r,i}\boldsymbol{\varepsilon}_{r,i}^{H} + \boldsymbol{v}_x^H\sum_{x=1}^{X}\boldsymbol{\varepsilon}_{\Phi,i,x}\boldsymbol{\varepsilon}_{\Phi,i,x}^{H}\boldsymbol{v}_x)
		= \sigma^{2}_{r,i} +\sum_{x=1}^{X} M_x\sigma^{2}_{\Phi,i,x}.
	\end{multline}
	This framework for imperfect CSI characterization offers insights into the impact of estimation errors on system performance and facilitates the development of robust communication strategies to mitigate their effects.

While the above framework assumes that imperfect channel estimates $\hat{{r}}_i$ and $\hat{\boldsymbol{\Phi}}_{i,x}$ are available at the BS, in practical RIS-assisted networks, acquiring this CSI introduces a substantial overhead. Specifically, the pilot-based training procedure is required to estimate both the direct user–BS links and the cascaded user–RIS–BS channels, which increases proportionally with the number of users, RISs, and reflecting elements. As a result, the effective time available for data and gradient transmission in each communication round is reduced, which may adversely affect the convergence rate and ultimate performance of FL. This highlights the importance of designing efficient CSI acquisition strategies, such as compressed sensing or machine learning–based channel prediction, which can mitigate the overhead while maintaining robust system performance.

{\bf{Imperfect SIC:}} Interference management represents a critical concern in hybrid networks comprising both AirFL and NOMA users, where all users are concurrently served over the entire bandwidth. Particularly, addressing interference from inter-type users is crucial. Employing SIC techniques, the BS sequentially decodes signals from strong users and then removes them to mitigate interference for the weak users \cite{10363658, 555555, 7959539}. To effectively manage interference in the proposed hybrid network, the BS employs SIC among NOMA users, while AirFL users adopt an analog superposition transmission scheme. In this setting, the local model updates transmitted by AirFL users are directly aggregated over the air without requiring individual decoding, enabling efficient over-the-air model aggregation. In contrast, NOMA users are decoded sequentially at the BS through SIC, where each user’s signal is detected after removing the stronger users’ signals.

To alleviate the interference from NOMA users to AirFL users, the SIC process is designed such that the BS decodes NOMA users first, treating the aggregated AirFL signal as background interference. The RIS phase-shift optimization is jointly configured to assist the NOMA users in satisfying their QoS requirements while simultaneously reducing the MSE in AirFL aggregation. Accordingly, without loss of generality, only the NOMA users are arranged in an ascending order of their effective channel gains to determine the SIC decoding order, expressed as
	\begin{equation} |\hat{{h}}_{K+1}^{(t)}|^{2} \leq |\hat{{h}}_{K+2}^{(t)}|^{2} \leq \ldots \leq |\hat{{h}}_{K+N}^{(t)}|^{2}, \quad \forall k \in \mathcal{K}.\end{equation}
This ordering assumption applies solely to NOMA users and serves as a tractable analytical simplification. It does not imply that the RIS can deterministically rank all composite channel gains, since the RIS provides only phase-based control.  Instead, its role is to enhance the statistical separation between users’ effective channels, thereby improving the probability of achieving a desirable SIC decoding sequence. Accordingly, the decoding order is dynamically updated at each time step based on the instantaneous effective channel gains, which depend on both channel conditions and RIS phase shifts.

While the RISs serves as a dynamic channel coordinator, enhancing the flexibility of SIC decoding order, imperfections in SIC execution may arise, affecting signal decoding. In practical scenarios, previously decoded signals may not be fully canceled, resulting in residual interference. Accordingly, the signal-to-interference-plus-noise ratio (SINR) of the $n$-th NOMA user at the BS under imperfect SIC is expressed as:
\begin{multline}
\gamma_n(t) = \\
{\frac{p_n^{(t)}\,|\bar{h}_n^{(t)}|^2}
{\underbrace{\sum\limits_{k=1}^{K} p_k^{(t)}\,|\bar{h}_k^{(t)}|^2}_{\substack{\text{AirFL interference}\\\text{(always present)}}}
+ \underbrace{\sum\limits_{j=K+1}^{n-1} p_j^{(t)}\,|\bar{h}_j^{(t)}|^2}_{\substack{\text{Uncanceled}\\\text{NOMA interference}}}
+ \underbrace{\epsilon_b \sum\limits_{j=n+1}^{K+N} p_j^{(t)}\,|\bar{h}_j^{(t)}|^2}_{\substack{\text{Residual NOMA}\\\text{interference after SIC}}}
+ \sigma^2}},
\end{multline}
where $\epsilon_b \in [0,1]$ denotes the imperfect SIC coefficient, capturing the residual interference from already-canceled NOMA users. This parameter captures practical SIC imperfections caused by channel estimation errors, hardware limitations, and error propagation during successive decoding. In this work, a small value of $\epsilon_b$ is adopted to reflect realistic receiver performance. An increase in $\epsilon_b$ leads to stronger residual interference, which degrades the SINR and consequently reduces the achievable data rates and overall system performance. Moreover, the SIC decoding order is determined based on the estimated effective channel gains, which reflects practical scenarios where only imperfect CSI is available at the receiver. In particular, $\epsilon_b = 0$ corresponds to ideal SIC, whereas $\epsilon_b = 1$ represents the absence of SIC cancellation. Here, the numerator represents the received power of the $n$-th NOMA user, while the denominator explicitly separates interferences from AirFL users, undecoded weaker NOMA users, and residual stronger NOMA users. The achievable uplink data rate for the $n$-th NOMA user is then:
	\begin{equation}
		R_n^{(t)} =B \log_2 (1 + \gamma_n^{(t)}),
	\end{equation}
	where $B$ denotes the bandwidth available at the BS. Thus, the overall uplink sum rate of the NOMA users is expressed as:
	\begin{equation}
		R_{\text{NOMA}}^{(t)} = \sum_{n=K+1}^{K+N} R_n^{(t)}.
	\end{equation}
Furthermore, due to the imperfections, the residual signal used in AirFL model aggregation can be represented as follows: 
	\begin{multline}
		\hat{{y}}^{(t)} = \sum_{k=1}^{K} (\hat{{h}}_k^{(t)}+\boldsymbol{\varepsilon}_{h,k}^{(t)}
		) \sqrt{p_k^{(t)}} s_k^{(t)} \\+ \sqrt{\epsilon_b} \sum_{n=K+1}^{K+N} (\hat{{h}}_n^{(t)}+\boldsymbol{\varepsilon}_{h,n}^{(t)}
		) \sqrt{p_n^{(t)}} s_n^{(t)} + \boldsymbol{z}_0^{(t)}.
	\end{multline}
	Subsequently, the BS applies a denoising factor, denoted as $\eta$, to the aggregated signal $\hat{{y}}^{(t)}$ to suppress noise. The reconstructed signal for AirFL users, represented as $\hat{s}^{(t)} = \frac{\hat{{y}}^{(t)}}{K\sqrt{\eta}}$, is then compared to the desired computation result from all AirFL users, denoted by $s^{(t)} = \frac{1}{K} \sum_{k=1}^{K} s_k^{(t)}$. It is important to note that $\eta$ alone does not fully compensate for heterogeneous channel gains, imperfect CSI, or residual SIC, and therefore does not guarantee fully unbiased aggregation. The residual misalignment arising from this simplification is explicitly reflected in the computation distortion.

The computation distortion of the recovered average gradient message with respect to the ideal average message $s^{(t)}$ under imperfect CSI and SIC scenario is quantified by the instantaneous MSE, defined as  
\begin{equation}
\text{MSE}(\hat{s}^{(t)}, s^{(t)}) = \mathbb{E}\big[|\hat{s}^{(t)} - s^{(t)}|^2\big],
\end{equation}
where the expectation is taken over the randomness of $s_i^{(t)}$, $\boldsymbol{\varepsilon}_{h,i}^{(t)}$, and $z_0^{(t)}$. Expanding the decomposition, we obtain:  
\begin{multline}
{\text{MSE}(\hat{s}^{(t)}, s^{(t)}) =
\underbrace{\frac{1}{K^2} \sum_{k=1}^{K} \Big|\frac{\hat{{h}}_k^{(t)} \sqrt{p_k^{(t)}}}{\sqrt{\eta}} - 1\Big|^2}_{\text{Signal misalignment error}} }\\ {
+ \underbrace{\frac{\epsilon_b}{K^2 \eta} \sum_{n=K+1}^{K+N} p_n^{(t)} \,|\hat{{h}}_n^{(t)}|^2}_{\text{SIC-related error}}
+ \underbrace{\frac{1}{K^2 \eta} \sum_{k=1}^{K} p_k^{(t)} \,\sigma_{h,k}^2}_{\text{CSI-related error}} }\\ { 
+ \underbrace{\frac{\epsilon_b}{K^2 \eta} \sum_{n=K+1}^{K+N} p_n^{(t)} \,\sigma_{h,n}^2}_{\text{SIC and CSI-related error}}
+ \underbrace{\frac{Q \sigma^2}{K^2 \eta}.}_{\text{Noise-induced error}}}
\end{multline}
Here, the first term corresponds to the signal misalignment error due to imperfect power and channel inversion; the second and fourth terms capture residual SIC effects with and without CSI errors; the third term reflects channel estimation error for AirFL users; and the last term accounts for the noise-induced error, scaled by the gradient dimension $Q$. 

The denoising factor $\eta$ plays a critical role in controlling the bias–variance tradeoff in AirFL aggregation. Specifically, $\eta$ acts as a normalization parameter for the received superimposed signal, balancing the effects of noise amplification and signal distortion. A smaller $\eta$ amplifies the aggregated signal, reducing the bias caused by channel inversion errors and signal misalignment, but increases the variance due to noise enhancement. In contrast, a larger $\eta$ suppresses noise, leading to lower variance at the cost of increased bias. Therefore, an appropriate choice of $\eta$ is essential to minimize the overall aggregation error.

\textbf{Remark 1:} In practical AirFL systems, propagation delays introduce phase offsets that affect coherent signal superposition. In this work, the RIS phase shifts are optimized to enhance the effective channel gain and reduce aggregation distortion; however, they are not explicitly designed for delay compensation (i.e., passive synchronization). The signal misalignment term in (16) captures the deviation from ideal coherent aggregation and thus implicitly accounts for residual phase offsets arising from propagation delays, imperfect channel inversion, and synchronization errors. This formulation provides a tractable way to model aggregation distortion without explicitly estimating per-user delays. 
\footnote{Future research directions include the development of delay-aware RIS phase optimization schemes that enable passive synchronization by compensating for propagation delay differences across users, thereby further improving coherent over-the-air aggregation performance.}.

\textbf{Remark 2:}  Achieving fully unbiased AirFL aggregation in practice would require per-user uplink scaling or channel inversion. Incorporating such uplink precoding is a natural extension of our framework and can be implemented in future work, while the current model provides a realistic estimate of the aggregation distortion under practical constraints.

	A thorough understanding of convergence behavior and the optimality gap in complex system settings is critical for effective resource allocation, performance optimization, and algorithm design. By analyzing convergence dynamics, valuable insights are gained into the efficiency and precision of iterative algorithms used in joint optimization tasks, such as power allocation, beamforming, and channel estimation. This understanding allows for the fine-tuning of algorithmic parameters and system configurations to meet the diverse needs of hybrid users, while also leveraging the additional degrees of freedom offered by RIS technology. Moreover, the evaluation of scalability and robustness of optimization algorithms across diverse operating conditions ensures the development of adaptive schemes capable of maintaining satisfactory convergence characteristics amidst evolving network dynamics and user scenarios.

	\section{Convergence Analysis and Problem Formulation}
	
	\subsection{Basic Assumption}
	
	To analyze the convergence properties of AirFL, we adopt the following assumptions, as outlined in \cite{9815289, 9606731, 9633251}.

	\textbf{Assumption 1 (L-Smooth):} The global loss function \( F(\boldsymbol{w}) \) is \( L \)-smooth, meaning there exists a constant \( L \) such that for any model parameters \( \boldsymbol{w} \) and \( \boldsymbol{v} \):
    	\begin{equation}
		F(\boldsymbol{w}) - F(\boldsymbol{v}) \leq \nabla F(\boldsymbol{v})^\top (\boldsymbol{w} - \boldsymbol{v}) + \frac{L}{2} \|\boldsymbol{w} - \boldsymbol{v}\|^2_2.
	\end{equation}

	\textbf{Assumption 2 (PL Inequality):} There exists a constant $\mu$ such that the Polyak-Lojasiewicz (PL) condition holds:
	\begin{equation}
		\left\| \nabla F(\boldsymbol{w}^{(t)}) \right\|^2_2 \geq 2\mu \left( F(\boldsymbol{w}^{(t)}) - F(\boldsymbol{w}^*) \right).
	\end{equation}

	\textbf{Assumption 3 (Gradient Bound):} The expected squared norm of each gradient is bounded:
	\begin{equation}
		\mathbb{E}\left[ \left\| \boldsymbol{g}_k^{(t)} \right\|_2^2 \right]\le \varpi^2.
	\end{equation}

	\textbf{Assumption 4 (Variance Bound):}The expected local gradients $\boldsymbol{g}_k$ at the $k$-th AirFL user are assumed to be independent and unbiased estimators of the global gradient $\boldsymbol{g} = \nabla F(\boldsymbol{w})$. Additionally, the variances of these gradients are bounded by specific limits, expressed as:
	\begin{align}
		\mathbb{E}[\boldsymbol{g}_k] &= \boldsymbol{g} \quad \forall k \in K, \\
		\mathbb{E}\left[(g_{k,i} - g_i)^2\right] &\leq \delta_i^2 \quad \forall k \in K, \ \forall i \in I, 
	\end{align}
	where $g_{k,i}$ and $g_i$ represent the $i$-th element of $\boldsymbol{g}_k$ and $\boldsymbol{g}$, respectively, and $\boldsymbol{\delta} = [\delta_1, \delta_2, \ldots, \delta_Q]$ represents a vector of non-negative variance bounds, ensuring controlled variance in gradient estimations.

	These assumptions collectively establish a robust theoretical foundation for conducting convergence analysis on AirFL, facilitating the development of tractable mathematical models to evaluate its performance under diverse conditions. Assumption 1, centered on Lipschitz smoothness, holds relevance across various prevalent loss functions, including linear regression, cross-entropy, logistic regression, and softmax classifiers. Assumption 2 comes from strong convexity. The parameters \( L \) and \( \mu \) in Assumptions 1 and 2 can be derived by computing the maximum and minimum eigenvalues of \( \nabla^2 F(w) \). Additionally, Assumptions 3 and 4 on gradient norms and variance bounds are standard in stochastic optimization literature, ensuring that gradients remain controlled, aiding convergence stability \cite{9815289, 9606731, 9633251}.

	\subsection{Convergence Analysis}
	
	Based on the reconstructed signal derived in (15), the BS is able to recover the local gradients ${\boldsymbol{g}_k^{(t)}}$ from the wireless signals ${s_k^{(t)}}$. Consequently, the received signal vector of interest at the BS is expressed as:
	\begin{multline}
		\hat{y}^{(t)} = \sum_{k=1}^{K} (\hat{h}_k^{(t)}+\boldsymbol{\varepsilon}_{h,k}^{(t)}
		) \sqrt{p_k^{(t)} }\boldsymbol{g}_k^{(t)} \\+ \sqrt{\epsilon_b} \sum_{n=K+1}^{K+N} (\hat{h}_n^{(t)}+\boldsymbol{\varepsilon}_{h,n}^{(t)}
		) \sqrt{p_n^{(t)}} s_n^{(t)} + z_0^{(t)}.
	\end{multline}
	Accordingly, the aggregated global gradient, as described in the before-mentioned section, can be expressed as:
	\begin{multline}
		\hat{\boldsymbol{g}}^{(t)} = \frac{\hat{\boldsymbol{y}}^{(t)}}{K\sqrt{\eta}} = \frac{1}{K\sqrt{\eta}} ( \sum_{k=1}^{K} (\hat{h}_k^{(t)}+\boldsymbol{\varepsilon}_{h,k}^{(t)}
		) \sqrt{p_k^{(t)}} \boldsymbol{g}_k^{(t)} \\+ \sqrt{\epsilon_b} \sum_{n=K+1}^{K+N} (\hat{h}_n^{(t)}+\boldsymbol{\varepsilon}_{h,n}^{(t)}
		) \sqrt{p_n^{(t)}} s_n^{(t)} + z_0^{(t)} ). 
	\end{multline}
	The global gradient estimation in (23) is significantly influenced by various communication factors, including transmit power, RIS configuration, noise, wireless channel conditions, and system imperfections. These factors contribute to the overall performance and accuracy of the gradient recovery process, which is crucial for the effectiveness of AirFL. To better understand the efficiency and convergence behavior of AirFL, it is important to investigate the total optimality gap, often referred to as the convergence upper bound. This gap provides insights into how well the learning algorithm converges over multiple communication rounds. In the following, we present a theoretical analysis that quantifies the learning performance of AirFL after $T$ communication rounds, offering a rigorous framework for evaluating its convergence behavior.

	{\bf{Theorem 1 (Optimality Gap):}}
	Let Assumptions 1 to 4 hold. Under the condition of a fixed learning rate that satisfies $0\le \lambda\le \frac{2}{2+L}$ and $\boldsymbol{\epsilon}^{(t)}=\hat{\boldsymbol{g}}^{(t)}-{\boldsymbol{g}}^{(t)}$ is defined as gradient aggregation error, after $T$ communication rounds, the expected optimality gap satisfies the following inequality:
	\begin{multline}
		\mathbb{E}\left[ F\left(\boldsymbol{w}^{(T+1)}\right) \right] - F\left(\boldsymbol{w}^*\right) \leq 
		(1-\lambda\mu)^T\left(  F(\boldsymbol{w}^{(1)}) - F(\boldsymbol{w}^{*}) \right)
		\\+\sum_{t=1}^{T}(1-\lambda\mu)^{(T-t)} \psi^{(t)},
	\end{multline}
	where, $\psi^{(t)}=\frac{L\lambda^2}{2}\| \boldsymbol{\delta
    }\|_2^2 +\frac{ 1+L^2\lambda^2 }{2}\| \mathbb{E}[\boldsymbol{\epsilon}^{(t)}] \|_2^2+\frac{L\lambda^2}{2}\mathbb{E}\left[\| \boldsymbol{\epsilon}^{(t)}\|_2^2\right]$. 
	
	\begin{IEEEproof} 
	Details of the proof are provided in Appendix.
	\end{IEEEproof}

	{\textbf{Remark 3:}}  The convergence behavior of this algorithm varies based on the nature of gradient aggregation. When aggregation is unbiased, meaning $\mathbb{E}[\boldsymbol{\epsilon}^{(t)}]=0$, the training model converges to the optimal point that minimizes the training loss as the number of iterations $T$ increases. In contrast, when aggregation is biased, the model may only converge to a neighborhood around the optimal point. However, ensuring unbiased aggregation can lead to slower convergence compared to biased aggregation, as ensuring unbiasedness typically entails increased MSE. Moreover, the sensitivity to aggregation errors becomes more pronounced in later communication rounds, since the discounting factor $(1-\lambda\mu)^{T-t}$ diminishes the impact of errors introduced in early rounds. Consequently, both the bias $\|\mathbb{E}[\boldsymbol{\epsilon}^{(t)}]\|_2^2$ and the variance $\mathbb{E}\!\left[\|\boldsymbol{\epsilon}^{(t)}\|_2^2\right]$ contribute more significantly to the optimality gap in later iterations.

    In the subsequent analysis, we rigorously evaluate the optimality gap of the AirFL framework by analyzing its dependency on transmission power control variables and denoising factors. Specifically, we express the aggregation error term $\boldsymbol{\epsilon}^{(t)}$ as a function of these parameters, thereby enabling a precise characterization of how power allocation and noise mitigation influence model convergence and accuracy. As detailed in Appendix A, the squared bias and MSE of the model estimates at each outer iteration $t$ are derived as follows:
	\begin{multline}
		\| \mathbb{E}[\boldsymbol{\epsilon}^{(t)}] \|_2^2=\left\|\frac{1}{K} \sum_{k=1}^{K}\left( \frac{\hat{h}_k^{(t)} \sqrt{p_k^{(t)}}}{\sqrt{\eta}}-1 \right) \mathbb{E}\left[ \boldsymbol{g}_k^{(t)} \right]\right\|_2^2 \\\le \frac{1}{K^2}\sum_{k=1}^{K}\left| \frac{\hat{h}_k^{(t)} \sqrt{p_k^{(t)}}}{\sqrt{\eta}}-1 \right|^2\varpi^2,
	\end{multline}
    and
	\begin{multline}
		\mathbb{E}\left[\| \boldsymbol{\epsilon}^{(t)}\|_2^2\right]=\mathbb{E}\left\|\frac{1}{K} \sum_{k=1}^{K}\left( \frac{(\hat{h}_k^{(t)}+\boldsymbol{\varepsilon}_{h,k}^{(t)}
			) \sqrt{p_k^{(t)}}}{\sqrt{\eta}}-1 \right)  \boldsymbol{g}_k^{(t)} \right\|_2^2\\+\mathbb{E}\left\| {\frac{\sqrt{\epsilon_b}}{K\sqrt{\eta}} \sum_{n=K+1}^{K+N} (\hat{h}_n^{(t)}+\boldsymbol{\varepsilon}_{h,n}^{(t)}
			) \sqrt{p_n^{(t)}} s_n^{(t)}} \right\|_2^2+\mathbb{E}\left\| {\frac{z_0^{(t)}}{K\sqrt{\eta}}} \right\|_2^2\\
		\le \frac{1}{K^2} \sum_{k=1}^{K}
		\left( \left| \frac{\hat{h}_k^{(t)} \sqrt{p_k^{(t)}}}{\sqrt{\eta}}-1 \right|^2+\frac{\sigma^{2}_{h,k}\ p_k^{(t)}}{\eta} \right) \varpi^2 \\+{\frac{\epsilon_b}{K^2{\eta}} \sum_{n=K+1}^{K+N} \left(
|\hat{h}_n^{(t)}|^2+\sigma_{h,n}^{2} \right) \ {p_n^{(t)}} }+\frac{Q\sigma^2}{K^2\eta},
	\end{multline}
	respectively, where these inequalities result from the combined use of Assumptions 3 and 4, along with the Cauchy’s inequality. Building upon Theorem 1, and by substituting (25) and (26) into (24), we derive the following proposition.
	{\bf{Proposition 1:}} For a learning rate $0\le \lambda\le \frac{2}{2+L}$, the expected optimality gap for AirFL satisfies:
\begin{multline}
    \mathbb{E}\!\left[ F\!\left(\boldsymbol{w}^{(T+1)}\right) \right] - F(\boldsymbol{w}^*) \leq 
(1-\lambda\mu)^T\!\left(  F(\boldsymbol{w}^{(1)}) - F(\boldsymbol{w}^{*}) \right) \\
+\sum_{t=1}^{T}(1-\lambda\mu)^{(T-t)} \Bigg( 
\frac{L\lambda^2}{2}\| \boldsymbol{\delta} \|_2^2 
+\\\frac{1}{K^2}\Bigg[ \frac{1+L^2\lambda^2}{2} 
\sum_{k=1}^{K}  
\left| \frac{\hat{h}_k^{(t)} \sqrt{p_k^{(t)}}}{\sqrt{\eta}} - 1 \right|^2 \varpi^2 \\+ \frac{L\lambda^2}{2}\sum_{k=1}^{K}\left( \left| \frac{\hat{h}_k^{(t)} \sqrt{p_k^{(t)}}}{\sqrt{\eta}} - 1 \right|^2+ \frac{\sigma^{2}_{h,k}\, p_k^{(t)}}{\eta}\  \right)\varpi^2 \\+\frac{L\lambda^2 \epsilon_b}{2K^2{\eta}} 
\sum_{n=K+1}^{K+N} \big( |\hat{h}_n^{(t)}|^2 + \sigma_{h,n}^{2} \big) p_n^{(t)} 
+ \frac{L\lambda^2 Q\sigma^2}{2K^2\eta} \Bigg] \Bigg)\\\triangleq {\Omega_T(\eta,p_i^{(t)},\theta_i^{(t)})}.
\end{multline}

\subsection{Problem Formulation}

Building on the insights presented, our focus shifts to improving the convergence rate by minimizing the identified optimality gap. To achieve this objective, we propose a joint optimization approach involving the denoising factor, user transmit power, and RIS configurations. This strategy is intended to accelerate convergence while maintaining communication quality. The optimization problem, subject to transmit power constraints, QoS requirements for NOMA users, and MSE tolerance for AirFL users, is formulated as:
\begin{equation}
	\begin{aligned}
		\boldsymbol{\mathcal{P}}_1: &\min \hspace{0.2cm} {\Omega_T(\eta,p_i^{(t)},\theta_i^{(t)})} && 
		\\
		&\text{s.t. } C_1:   B\log_2{(1+\gamma_n^{(t)})}\ge R_n^{\text{min}}, \ \ \forall n \in \mathcal{N}, \\
		& \ \ \ C_2: \text{MSE}(\hat{s}^{(t)}, s^{(t)})\le \epsilon_0,  \\
		& \ \ \ C_3:0 \leq \theta_m^{(t)} < 2\pi,  \ \ \ \ \ \forall m \in \mathcal{M}, \\
		& \ \ \ C_4:\eta > 0,\\		  
		& \ \ \ C_5: 0 \;\le\; p_i^{(t)} \;\le\; P_{\mathrm{max}}, \qquad \forall i\in\mathcal{I},\\		  
		& \ \ \ C_6:  (10),   
	\end{aligned}
\end{equation}
where \(C_1\) refers to the QoS requirement at the $n$-th NOMA user; \(C_2\) refers to the maximum permissible MSE for AirFL users, with $\epsilon_0 > 0$; \(C_3\) refers to the range of phase shifts for the RIS elements; \(C_4\) applies to the range of the denoising factor; \(C_5\) refers to the maximum transmit power constraint of each users, and \(C_6\) refers to the SIC decoding order.

Solving problem $\boldsymbol{\mathcal{P}}_1$ directly is challenging due to its non-convex objective function and the stochastic nature of the dynamic environment, particularly in massive mobile IoT scenarios. To address this challenge, we explicitly leverage the recursive characterization of the optimality gap provided in Appendix of Theorem~1:
\begin{equation}
	\Theta^{(t+1)}\le (1-\lambda\mu)\Theta^{(t)}+\psi^{(t)},
\end{equation}
where $0<\lambda\mu<1$ guarantees geometric contraction of the first term. This relation indicates that $\psi^{(t)}$ acts as the additive error term controlling the rate of convergence. By minimizing $\psi^{(t)}$ at each training round, we iteratively suppress the residual error and ensure that the cumulative gap $\Omega_T = \sum_{t=1}^{T} \Theta^{(t)}$ is upper-bounded by a convergent series. Although this transformation is locally greedy, it tracks the solution trajectory of $\boldsymbol{\mathcal{P}}_1$ and asymptotically approaches the global optimum as $t \to \infty$.

Moreover, we emphasize that solving $\boldsymbol{\mathcal{P}}_1$ in a fully offline manner is computationally prohibitive in dynamic IoT networks because it requires global knowledge of all $T$ training rounds. Reformulating the problem into an online version enables a tractable, adaptive solution that responds to instantaneous channel variations and user mobility, offering a favorable trade-off between optimality and complexity.

Based on this reasoning, $\boldsymbol{\mathcal{P}}_1$ is reformulated into the following online problem:
\begin{equation}
	\begin{aligned}
	\boldsymbol{\mathcal{P}}_2: &\min_{\ \ \eta,p_i^{(t)},\theta_i^{(t)}}\hspace{0.2cm} \psi^{(t)} \\
		\ \ \ \ &\ \text{s.t. } C_1, C_2, C_3, C_4,C_5.
	\end{aligned}
\end{equation}
This reformulation thus guarantees that the per-round solution progressively reduces the global gap while remaining computationally feasible, making it well-suited for large-scale IoT scenarios where real-time adaptation is critical.

\textbf{Remark 4:} It should be noted that the per-round minimization of $\psi^{(t)}$ in $\boldsymbol{\mathcal{P}}_2$ represents a greedy optimization strategy. While this approach does not strictly guarantee the globally optimal convergence rate over $T$ rounds, the recursive inequality in (29) ensures that the optimality gap evolves as a contractive process with bounded additive error. In practice, by minimizing $\psi^{(t)}$ at each round, the cumulative gap $\Omega_T$ remains bounded and converges to a neighborhood of the global optimum. This provides a practical trade-off between tractability and convergence performance in dynamic wireless environments.

Nevertheless, the non-convex nature of $\boldsymbol{\mathcal{P}}_2$, coupled with the heterogeneity of dynamic environments, renders conventional optimization methods impractical for real-time operation. To address this challenge, we propose a learning-based framework that leverages the temporal structure of the problem. Specifically, we adopt an LSTM-DDPG architecture to model long-term dependencies in system dynamics and learn efficient online policies without requiring knowledge of future rounds.

	\section{Proposed DRL Algorithm}

	We reformulate the optimization problem as an MDP, allowing us to leverage DRL through the LSTM-DDPG algorithm. 
	
	An MDP serves as a formal mathematical model for describing decision-making problems, in which an agent makes decisions by interacting with an environment at discrete time steps. The aim of the MDP is to accomplish a given task by maximizing its cumulative rewards, which can be formally represented by the tuple $(\mathcal{S}, \mathcal{A}, \mathcal{P}, \mathcal{R})$. In this formulation, \( \mathcal{S} \) denotes a set of possible states the environment can be in, providing the context for the agent's decisions, and \( \mathcal{A} \) represents a set of actions available to the agent, which defines its decision-making choices. The state transition function \( \mathcal{P} \) maps each state-action pair to a probability distribution over subsequent states, allowing the agent to estimate the likely outcomes of its actions. The reward function \( \mathcal{R} \) assigns a reward to the agent based on the current state, the chosen action, and the resulting state, facilitating the agent's learning by reinforcing favorable outcomes.

	At each time step \( t \), the agent observes the state \( \boldsymbol{s}^{[t]} \in S \) and chooses an action \( \boldsymbol{a}^{[t]} \in A \) in accordance with a policy \( \pi_{\phi} \), parameterized by \( \phi \). Following the action selection, the agent receives an immediate reward \( r^{[t]} \in R \) and transitions to a new state, thereby progressing within the environment. The primary objective for the agent in an MDP is to determine the optimal policy that maximizes the expected long-term cumulative rewards, which can be expressed as:
	\begin{equation} 
		R = \sum_{t=0}^{T} \gamma^{(t)} r(\boldsymbol{s}^{(t)}, \boldsymbol{a}^{(t)}), 
	\end{equation}
	where \( \gamma \in [0, 1] \) is the discount factor that balances short-term and long-term rewards, enabling the agent to avoid short-sighted behavior and consider future rewards.
	
	To approximate the optimal solution, which is often infeasible to obtain analytically, we leverage DRL techniques. Specifically, a LSTM network is used instead of a traditional feedforward network to capture complex, long-term dependencies in decision-making. The proposed system can thus be represented as an MDP with well-defined states, actions, and reward functions as:
	
	\textbf{State Space:} At each time slot \( t \), the state space includes the channel gain estimates of all links, represented by:
	\begin{equation} 
		\boldsymbol{s}^{(t)} = [{r}_{i}^{(t)}, \boldsymbol{G}^{(t)}, {h}_i^{(t)}], \quad \forall i \in \mathcal{I}. 
		\label{eq}
	\end{equation}
	
	\textbf{Action Space:} The action space consists of the transmit power vector, \( \boldsymbol{P} = [p_1, \dots, p_{N+K}]\), the phase shift vector of each RIS, \( \boldsymbol{\theta} = [\boldsymbol{\theta}_1, \dots, \boldsymbol{\theta}_X] \), where \( \boldsymbol{\theta}_x \in \mathbb{R}^{M_x} \), and the denoising factor \( \eta \). Consequently, the action space is represented as:
	\begin{equation} 
		\boldsymbol{a}^{(t)} = [\boldsymbol{P}^{(t)}, \boldsymbol{\theta}^{(t)}, \eta^{(t)}]. 
	\end{equation}
   For each selected action, the effective channel gains are computed, and the corresponding NOMA decoding order is updated based on (10). The resulting SINR values are then used to evaluate the reward, making the decoding order an implicit function of the agent’s action.
	
	\textbf{Reward Function:} To guide the agent toward optimizing the objective function under given constraints, the reward function at each time slot is expressed as:
	\begin{equation} 
		r(\boldsymbol{s}^{(t)}, \boldsymbol{a}^{(t)}) = -\psi^{(t)} + r_m  \chi_m^{(t)} + r_s  \chi_s^{(t)}, 
	\end{equation}
	where \( \chi_m^{(t)} \) and \( r_m \) are penalty indicators and factors associated with the MSE constraint \( C_1 \). Specifically, \( \chi_m^{(t)} = 1 \) if the MSE exceeds a threshold, and \( \chi_m^{(t)} = 0 \) otherwise. Similarly, \( \chi_s^{(t)} \) and \( r_s \) represent the penalty indicators and factors for the SINR constraint \( C_2 \), with \( \chi_s^{(t)} = 1 \) if the SINR falls below a specified threshold, and \( \chi_s^{(t)} = 0 \) otherwise.
	The penalty factors \( r_m \) and \( r_s \) are tuned through trial and error to balance exploration and exploitation, ensuring optimal convergence in simulations. These factors play a critical role in shaping the agent’s learning behavior. Intuitively, they determine how strongly the agent is discouraged from violating the feasibility constraints. If these values are too small, the agent may overlook constraint violations in favor of short-term performance, which can yield infeasible or unstable solutions. On the other hand, if they are set excessively large, the penalties may dominate the reward signal, suppressing exploration and destabilizing the training process. Hence, the calibration of these parameters is essential to strike a balance between constraint enforcement, convergence stability, and effective exploration.  
    By leveraging these tailored states, actions, and rewards, our proposed MDP model provides a robust framework for optimizing long-term objectives in dynamic wireless environments.

	We employ DRL algorithms to address the MDP problem. Model-free value-based DRL algorithms such as DQN are not suitable for the considered problem due to the continuous nature of the action space, which includes transmit power allocation, RIS phase shifts, and the denoising factor. Applying DQN would require discretization of these variables, resulting in a significant increase in action space dimensionality and a loss of control precision. This would lead to higher computational complexity and degraded performance. Consequently, we utilize policy gradient-based reinforcement learning methods, which optimize the policy by leveraging the gradient of the expected reward. To ensure stable and efficient training, both state and action variables are normalized prior to being processed by the neural networks. This normalization mitigates scale imbalance among input features and improves convergence behavior. The DDPG algorithm, an off-policy actor-critic method, is well-suited for continuous action spaces. However, traditional deep neural networks (DNNs) typically used in DDPG fail to capture temporal patterns in environmental dynamics \cite{sheikhzadeh2021ai}. To enhance the model's capacity for recognizing these temporal dynamics, we propose an LSTM-DDPG approach. This architecture incorporates LSTM networks within the DDPG framework, allowing for more effective learning of temporal dependencies and adaptive responses to dynamic environmental changes.
The environment considered in this work is not memoryless across time slots. Specifically, the channel follows a block-fading model with temporal evolution, where large-scale components vary gradually over time, while small-scale fading is independently generated at each time slot. In addition, the sequential nature of resource allocation decisions introduces temporal dependency in the state transitions. Therefore, the system exhibits partial temporal correlation, which justifies the use of LSTM to effectively capture such dependencies.

	The proposed LSTM-DDPG architecture consists of four key neural networks: an actor network (policy network), a critic network (Q-network), and their respective target networks. The actor network, denoted by $\pi_{\phi}$ with parameters $\phi$, outputs actions $\boldsymbol{a}^{(t)} = \pi_{\phi}(\boldsymbol{s}^{(t)}) + \xi$ based on the current state $\boldsymbol{s}^{(t)}$, where $\xi$ is an added noise term to facilitate exploration. The critic network, parameterized by $\theta_Q$, estimates Q-values $Q_{\theta_Q}(\boldsymbol{s}^{(t)}, \boldsymbol{a}^{(t)})$ for state-action pairs. Additionally, a target actor network and a target critic network serve as delayed versions of the actor and critic, respectively, for stability in training. The integration of LSTM layers into the actor-critic structure enables the model to capture and exploit long-term temporal correlations in state transitions, thereby improving performance in dynamic environments.
	
	The proposed algorithm aims to optimize a policy that maximizes the expected cumulative reward from a given state $\boldsymbol{s}^{(t)}$ and action $\boldsymbol{a}^{(t)}$, expressed as:
	\begin{equation}
		\pi^* = \arg \max_{\pi} \mathbb{E}_{\boldsymbol{s}^{(t)}, \boldsymbol{a}^{(t)}} \left[ \sum_{t=0}^{\infty} r(\boldsymbol{s}^{(t)}, \boldsymbol{a}^{(t)}) \right].
	\end{equation}
This is achieved by optimizing the actor network using a policy gradient method to maximize the objective function \( J(\phi) \):
	\begin{multline}
		\nabla_{\phi_{\pi}} J(\phi_{\pi}) =\\ \mathbb{E} \left[ \nabla_{\boldsymbol{a}^{(t)}} Q_{\theta_{Q}}(\boldsymbol{s}^{(t)}, \boldsymbol{a}^{(t)}) \bigg|_{\boldsymbol{a}^{(t)}=\pi_{\phi_{\pi}}(\boldsymbol{s}^{(t)})} \nabla_{\phi_{\pi}} \pi_{\phi_{\pi}}(\boldsymbol{s}^{(t)}) \right].
	\end{multline}
	Concurrently, the critic network is trained to minimize the loss function concerning the target value \( Y_t \), defined as:
	\begin{equation}
		Y^{(t)} = r^{(t)} + \gamma Q_{{\theta'_{Q}}}(s^{(t+1)}, \pi_{\phi'_{\pi}}(s^{(t+1)}) + {\xi}).
		\label{eq:target}
	\end{equation}
    The LSTM-DDPG method is outlined in Algorithm 1.

	\RestyleAlgo{ruled}
	\SetNlSty{textbf}{}{:}
	\begin{algorithm}
		
		\SetAlgoLined
		\textbf{Initialization:} Initialize the experience replay buffer \( M_r \), set the mini-batch size \( H \), and assign random weights to the actor network \( \pi_{\phi_{\pi}} \) and the critic network \( Q_{\theta_Q} \). Create target networks by copying the current networks: \( \theta'_Q \gets \theta_Q \) and \( \phi'_{\pi} \gets \phi_{\pi} \).
		\\        
		\textbf{Define:} Maximum episodes \( E \) and episode length \( T_s \).
		\\
		\For{each episode $e \in \{1, \ldots, E\}$}{%
Initialize the starting state $\boldsymbol{s}^{(0)}$ and exploration noise $\boldsymbol{\xi}$;
			\\
			\For{$t = 1$ \KwTo $T_s$}{%
				Observe the current state $\boldsymbol{s}^{(t)}$;
				\\
				Compute the action $\boldsymbol{a}^{(t)} = \pi_{\phi_{\pi}}(\boldsymbol{s}^{(t)}) + \boldsymbol{\xi}$ using the actor network and apply any necessary reshaping;
				\\
				Retrieve the reward $\boldsymbol{r}^{(t)}$ based on (34);
				\\
				Observe the next state $\boldsymbol{s}^{(t+1)}$;
				\\
				Store the transition $(\boldsymbol{s}^{(t)}, \boldsymbol{a}^{(t)}, \boldsymbol{r}^{(t)})$ in the replay buffer $M_r$;
			}
			Randomly select a mini-batch of $H$ transitions from $M$;
			\\
			Calculate the target value $Y^{(t)}$ based on (37);
			\\
			Update both the actor and critic networks using the Adam optimization algorithm;
			\\
			Perform a soft update on the target networks with a soft update coefficient \(\tau \in [0, 1]\) as follows:
			\[
			\phi'_{\pi} \leftarrow \tau \phi_{\pi} + (1 - \tau) \phi'_{\pi}, \quad
			\theta'_Q \leftarrow \tau \theta_Q + (1 - \tau) \theta'_Q
			\]
		}
		\caption{LSTM-DDPG Algorithm}
		\label{algorithm1}
	\end{algorithm}

	\subsection{Computational Complexity Analysis}
	
	The computational complexity of a DRL network involves both the action selection and training processes \cite{gharehgoli2023ai,sheikhzadeh2021ai}. This network architecture comprises one actor network and one critic network with a fixed number of hidden layers, denoted by $\mathcal{U}$, each with the same number of neurons per hidden layer, denoted by $\mathcal{P}$. The complexity related to action selection, which refers to generating network output for a given input, is quantified by the cumulative product of the sizes of consecutive layers \cite{gharehgoli2023ai,sheikhzadeh2021ai}. Owing to the utilization of previous trajectories $\mathcal{J}$ for action selection, the production of the consequence layers in the actor network is as follows: 
	\(\mathcal{J} \times (|S| + |A|) \times \mathcal{P}\) for the input and first layer, 
	\(\mathcal{P}^2\) for the successive hidden layers, and 
	\(\mathcal{P} \times |A|\) for the output layer.
	
	Similarly, for the critic network, the production of the consequence layers is \(\mathcal{J} \times (|S| + 2 \times |A|) \times \mathcal{P}\) for the input and first layer, \(\mathcal{P}^2\) for the successive hidden layers, and 
	\(\mathcal{P} \times |A|\) for the final connection. Here, \(|S|\) and \(|A|\) denote the dimensions of the agent state and action spaces, respectively. Thus, the computational complexity of action selection for proposed method is \(\mathcal{O}(\mathcal{P}^2)\). 
	In the training process of LSTM-DDPG networks, the complexity is associated with the total number of edges, which can be expressed as \( I \times C + C^2 + C \times O \), where \( I \) denotes the number of inputs, \( C \) represents the number of neurons, and \( O \) indicates the number of outputs \cite{sheikhzadeh2021ai}. For the actor and critic networks, the edge counts are \((H |S| \mathcal{P} + H \mathcal{P}^2 + H \mathcal{P} |A|)\) and \((H (|S| + |A|) \mathcal{P} + H \mathcal{P}^2 + H \mathcal{P})\), respectively, where \( H \) represents the batch size. As a result, the complexity of the training process is \(\mathcal{O}(H \mathcal{P}^2)\). Using the same method, the computational complexities of both SAC and DDPG are \( \mathcal{O} \left( (\sum_{\mathcal{N}=1}^{\mathcal{N}_l} C_{\mathcal{N}} C_{\mathcal{N}-1}) H \mathcal{N}_e \right) \), where \( \mathcal{N}_l \) is the number of total layers, \( C_{\mathcal{N}} \) is the number of neurons in layer \( \mathcal{N} \), and \( \mathcal{N}_e \) is the number of total episodes \cite{gharehgoli2023ai}. 
    
Compared to conventional DRL methods, the proposed LSTM-DDPG introduces additional computational overhead due to its recurrent structure. However, this overhead primarily affects constant factors, while the overall computational complexity remains on the same order as conventional actor–critic methods. In practice, the training phase is performed offline, whereas the online inference stage only requires forward propagation through the trained networks, resulting in comparable latency for real-time decision-making. Moreover, by exploiting temporal correlations in dynamic environments, LSTM-DDPG achieves faster convergence and improved stability, which can reduce the total training time. Therefore, the proposed approach offers a favorable trade-off between computational cost and performance gain, making it suitable for practical deployment in RIS-assisted wireless systems.

	\section{Simulation Results}

  This section details the simulation setup and evaluation of the proposed LSTM-DDPG algorithm for efficient learning and resource allocation in a multi-RIS-assisted hybrid network. Its performance is compared with DDPG, SAC, and A2C to assess efficiency and adaptability. Additionally, the algorithm's learning capability is evaluated using the MNIST dataset to validate its robustness across different tasks.

	\subsection{Environment Settings} 
	
	We consider a multi-RIS-assisted hybrid network comprising 18 hybrid users and a single BS. Among these users, $K = 14$ are designated as AirFL users, and $N = 4$ as NOMA users. All users are uniformly and randomly positioned within a circular area of radius 100 m, centered at the BS located at $(0, 0, 20)$. The reference path loss is set to $\varrho_0 = 10^{-3}$, and a Rician fading factor of $\kappa = 2$ is applied. Each user $i$ operates with a power budget of $P_\text{max} = 10$ dBm. For each NOMA user $n$, the minimum rate requirement is specified as $R^{\text{min}}_n =1 \text{Mbps}$, $\forall n \in [K+1, K+N]$. Additionally, an aggregation error tolerance of $\epsilon_0 = 0.01$ is imposed for AirFL users. The number of RISs is set to 3, each containing $M = 20$ reflecting elements.

	Additionally, the algorithms under consideration use a learning rate of 0.0005, a replay buffer size of 600,000 samples, and a batch size of 64. The soft update parameter is set to 0.001, allowing for gradual policy updates to maintain stability, while the discount factor is chosen as 0.8 to balance immediate and future rewards. A maximum of 2200 episodes, each containing 100 time slots, is permitted. The environment resets either when the time slot limit is reached or when $R_{\text{min},n} < R_{\text{min}}$. Table \ref{sim} provides a summary of the simulation parameters.

	\begin{table}[htbp]
		\caption{System Parameters \cite{gharehgoli2023ai,9798757, 11342412, sheikhzadeh2021ai, 
 yuan2020intelligent}}
		\label{sim}
		\centering
		\begin{tabular}{|l|l|l|}
			\hline
			\textbf{Parameter} & \textbf{Description} & \textbf{Value} \\
			\hline
			$B$ & Bandwidth & 1 MHz \\
			\hline
			$K$ & Number of AirFL user & 14 \\
			\hline
			$N$ & Number of NOMA users & 4 \\
			\hline
			$X$ & Number of RISs & 3 \\
            \hline
			$M$ & Number of RIS elements & 20 \\
			\hline
			$\sigma^2$ & Total noise power over $B$ & -114 dBm \\
			\hline
			$P_{\text{max}}$ &\makecell[l]{ Maximum power constraint } & 10 dBm \\
			\hline
			$\epsilon_b$ & \makecell[l]{The imperfect SIC coefficient} & 0.04 \\
			\hline
			$\epsilon_{h}$ & \makecell[l]{The imperfect CSI component} & 0.02 \\
			\hline
			$h$ & Channel gain & Rayleigh Fading \\
			\hline
			$\eta_{r}$ & \makecell[l]{Path loss exponent for RIS-assisted \\channel} & 2.2 \\
			\hline
			$\eta_{d}$ & \makecell[l]{Path loss exponent for direct channel} & 3.5  \\
			\hline
			$\varrho_0$ & Path loss at reference distance & $10^{-3}$ \\
			\hline
			$\kappa	$ & Rician factor & 2  \\
			\hline
			$\bar{d}$ & The element separation distance & 0.5 $\bar{\lambda}$\\
			\hline
			$R_n^{min}$ & \makecell[l]{The minimum data rate required for\\ the $n$-th NOMA user} & 1 Mbps\\
			\hline
			$\epsilon_0$ & \makecell[l]{The tolerance for aggregation errors\\ among AirFL users} & 0.01\\
			\hline
			$r_m$ & Penalty factor for MSE requirement & -1 \\
			\hline
			$r_s$ & Penalty factor for bit rate requirement & -1 \\
			\hline
			$E$ & Number of episodes & 2200 \\
			\hline
			$T_s$ & Number of time slots & 100 \\
			\hline
			$ H$ & Batch size & 64 \\
			\hline
			$\tau$ & \makecell[l]{Target network update period} & 0.001 \\
			\hline
			$\mathcal{U}$ & Number of hidden layers & 2 \\
			\hline
			$\mathcal{P}$ & \makecell[l]{Number of neurons per hidden layer} & 512 \\
			\hline
			$-$ & \makecell[l]{Activation function in hidden layers} & ReLU  \\
			\hline
			$-$ & \makecell[l]{Activation function in output layer} & tanh \\
			\hline
			$-$ & Actor network learning rate & 0.0005 \\
			\hline
			$-$ & Critic network learning rate & 0.0001 \\
			\hline
			$\gamma$ & Discount factor & 0.8\\
			\hline
			$M_r$ & Replay buffer size & 600000 \\
			\hline
			
		\end{tabular}
		\label{tab:sample}
	\end{table}

     To assess the effectiveness of the proposed scheme, we compare its performance with the following benchmarks:
	
	\begin{itemize}
		
		\item \textbf{Hybrid Network without RIS}: This configuration includes a single BS and $I$ devices within a hybrid network, where communication and AirFL users transmit over a single MAC to compute specified functions.
		
		\item \textbf{Hybrid Network with Single-RIS}: In this scenario, we enhance the baseline configuration by introducing a single central RIS positioned at coordinates (50, 0, 20). This RIS aims to facilitate efficient model uploading from devices to the BS. To ensure a fair comparison with the proposed multi-RIS setup, the number of reflecting elements in this single-RIS configuration is standardized to $X \times M$.

\item \textbf{Hybrid Network with Multi-RIS}: In this configuration, three RISs are strategically deployed to enhance the communication links between users and the BS. The RISs are positioned at fixed coordinates to provide effective coverage and create multiple reflection paths: RIS~1 at \((50, 50, 10)\) meters, RIS~2 at \((-50, 50, 10)\) meters, and RIS~3 at \((0, -50, 10)\) meters. Each RIS is mounted at a height of 10 meters, reflecting realistic deployments on building facades or urban infrastructure.

		\item \textbf{Hybrid Network with Random Phase Shifts}:
		In this model, a single RIS is employed, with each reflecting element's phase shift independently and uniformly assigned within the interval \([0, 2\pi)\). This configuration allows us to isolate the impact of optimized phase shifts by comparing it against a network where the RIS phase shifts are not controlled by the LSTM-DDPG algorithm. In this configuration, the LSTM-DDPG algorithm is applied only to optimize other system parameters, excluding the random phase shifts.
		
	\end{itemize}
	
	In the upcoming experiments, the average reward at episode $e$ is calculated using the formula \({R_{avg}(e)} = \frac{1}{100} \sum_{i=e-100}^{e} R(i)\), where \(R(i)\) represents the mean reward of episode \(i\).

		\subsection{Performance Evaluation}
	This subsection presents a comparison of the proposed algorithm with several DRL methods, analyzing the impact of key parameters such as the number of AirFL users, power budgets, CSI and SIC imperfections, and RIS reflecting elements. The performance is benchmarked under various algorithms, highlighting the efficacy of the proposed method. Furthermore, we evaluate the ability of the LSTM-DDPG algorithm to manage FL tasks using the MNIST dataset, demonstrating its superior performance in these scenarios.

\begin{figure}[ht]
		\centering
		\includegraphics[width=3.1in, height=1.5in]{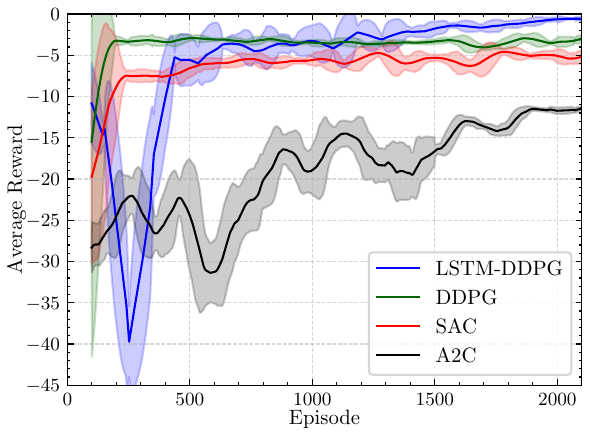}
		\caption{Comparison of DRL algorithms based on the average reward across training episodes.}
		\label{fig:episode}
	\end{figure}

	\textit{1) DRL comparison:}
	Fig.~\ref{fig:episode} illustrates the average rewards (solid lines) along with the standard deviations (shaded areas) over different episodes for various DRL algorithms. As training progresses, all schemes demonstrate an upward trend in average rewards per episode, confirming that the agents are learning to reduce the optimality gap effectively. Notably, the LSTM-DDPG algorithm converges faster, approximately after 1500 episodes, and achieves higher average rewards with lower variance compared to standard DRL algorithms. It is worth noting that LSTM-DDPG exhibits a transient sharp decrease in average reward during early training stages (around episode 300). This behavior can be attributed to the increased exploration induced by the LSTM’s memory effect and the stochastic nature of policy updates in dynamic environments. Specifically, as the LSTM-DDPG agent leverages historical state information for decision-making, it tends to explore diverse action trajectories more aggressively, which occasionally results in suboptimal rewards before the policy stabilizes. However, this exploration phase is crucial for escaping local optima and achieving superior long-term performance. After sufficient learning, the algorithm successfully leverages temporal dependencies to enhance resource allocation decisions, resulting in more stable and improved rewards in later episodes. This observation underscores the superior performance and enhanced stability of the LSTM-DDPG algorithm in dynamic environments.

\begin{figure}[t]
		\centering
		\includegraphics[width=3.1in, height=1.5in]{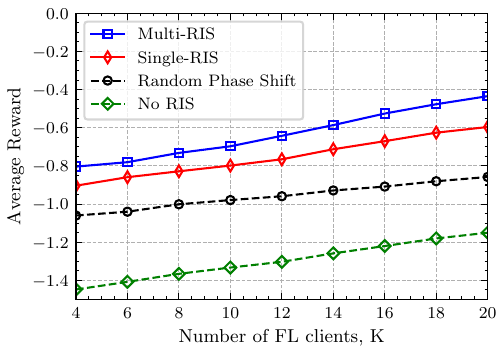}
		\label{fig:power}
		\caption{Average reward versus different number of AirFL users.}
		\label{fig:comparison_b1}
	\end{figure}

	\textit{2) Assessment of different numbers of AirFL users:} Fig.~\ref{fig:comparison_b1} presents the performance of the RIS-assisted hybrid network as the number of AirFL users increases, while keeping the number of NOMA users constant. The results reveal a steady improvement in the average reward across all evaluated schemes as the count of AirFL users grows. Notably, configurations with multiple RIS units consistently outperform other benchmarks, highlighting their role in mitigating interference and enhancing function aggregation. Although incorporating multiple RIS units may incur additional production costs, their ability to support greater user loads while maintaining high performance demonstrates the potential of RIS technology to meet the demands of dense, hybrid networks.

\begin{figure}[t]
		\centering
		\includegraphics[width=3.1in, height=1.5in]{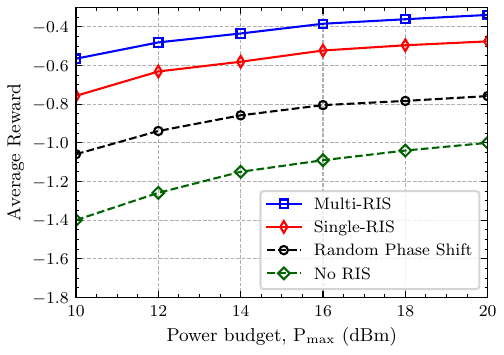}
		\label{fig:power}
		\caption{Average reward versus power budgets of users.}
		\label{fig:comparison_b}
	\end{figure}

	\textit{3) Assessment of different power budget:} Fig.\ref{fig:comparison_b} presents the average reward versus the maximum transmit power at the users. Notably, the average reward for all schemes increases with \( P^{\text{max}} \), with the proposed scheme outperforming the other methods. The power budget has a notable effect on the ability of users to upload and aggregate their data. In a scenario with limited transmit power, users face more difficulty in maintaining reliable connections, particularly when the channel conditions are not ideal. This results in lower average rewards, as the function aggregation process becomes less efficient. Conversely, higher power budgets enable more effective data transmission and aggregation, resulting in improved rewards.
	
\begin{figure}[t]
		\centering
		\includegraphics[width=3.1in, height=1.5in]{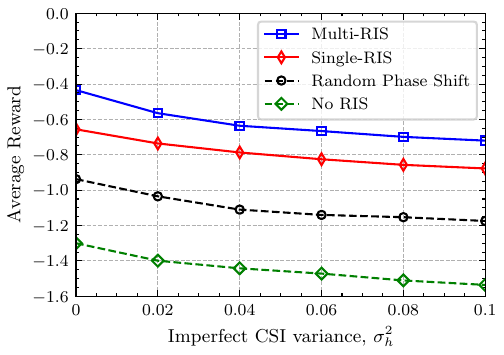}
		\label{fig:NUMRIS1}
		\caption{Average reward versus the CSI imperfection.}
		\label{fig:comparison_cc1}
	\end{figure}

\begin{figure}[t]
		\centering
		\includegraphics[width=3.1in, height=1.5in]{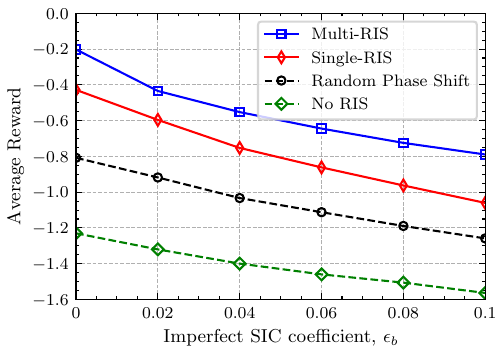}
		\label{fig:NUMRIS2}
		\caption{Average reward versus the SIC imperfection.}
		\label{fig:comparison_cc2}
	\end{figure}

	\textit{4) Assessment of the imperfect CSI and SIC:} Figs. \ref{fig:comparison_cc1} and \ref{fig:comparison_cc2} show that increasing CSI and SIC imperfections consistently reduce the average reward. In Fig. \ref{fig:comparison_cc1}, this decrease can be attributed to reduced channel estimation accuracy, which undermines the effectiveness of resource allocation and beamforming, leading to suboptimal network performance and heightened interference for both NOMA and AirFL users. In Fig. \ref{fig:comparison_cc2}, the results align with expectations, showing that imperfect SIC introduces more residual interference than ideal SIC, thereby further reducing the average reward. This increased interference particularly affects NOMA and AirFL users, as the higher uncertainty exacerbates signal decoding challenges and amplifies interference, ultimately degrading overall network performance.

\begin{figure}[t]
		\centering
		\includegraphics[width=3.1in, height=1.5in]{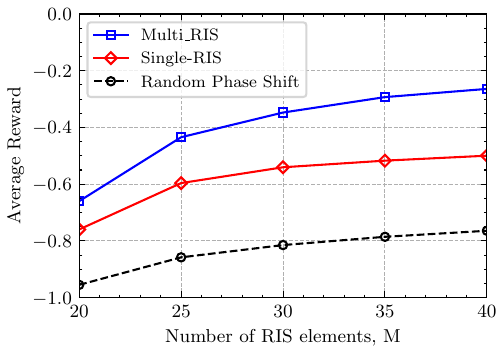}
		\label{fig:NUMRIS4}
		\caption{Average reward versus the number of reflecting elements at RIS.
        }
		\label{fig:comparison_c2}
	\end{figure}

	\textit{5) Assessment of the number of reflecting elements:} Fig. \ref{fig:comparison_c2} illustrates the relationship between the average reward and the number of reflecting elements $M$ at the RIS. The results indicate a positive trend, where the average reward consistently increases as $M$ grows, across all evaluated cases. This outcome underscores the effectiveness of deploying additional reflecting elements at the RIS, as more elements enhance signal reflection capabilities, improve channel conditions, and facilitate more efficient resource allocation.

	\textit{6) Assessment of performance in real FL tasks:}
	To effectively assess the performance of our proposed method for handling FL tasks, we conducted training on image classification models using the MNIST dataset. It is worth noting that the use of the MNIST dataset enables validation of the proposed algorithm using real data, capturing practical data heterogeneity across edge devices. This complements the wireless system simulations and demonstrates the effectiveness of the proposed approach in realistic learning scenarios. The dataset was initially split into training \(90\%\) and test \(10\%\) sets. The training data was further split into 14 partitions, each allocated to a different client without replacement to simulate data heterogeneity across clients. Each client trained a local model based on a feedforward neural network architecture, consisting of an input layer with 200 neurons and ReLU activation, followed by a hidden layer with 200 neurons and ReLU activation. The output layer includes a neuron for each class in the classification task, utilizing a softmax activation function. As shown in Fig. 8, the LSTM-DDPG approach outperforms both random phase shift and no-RIS configurations, achieving lower training loss and higher test accuracy. It can also be observed that the proposed algorithm converges rapidly and reaches stable performance within a limited number of iterations, and further increasing the number of iterations does not yield significant additional improvement, confirming the convergence stability of the proposed approach.

	\begin{figure*}
		\centering{		\includegraphics[width=6.2in, height=1.5in]{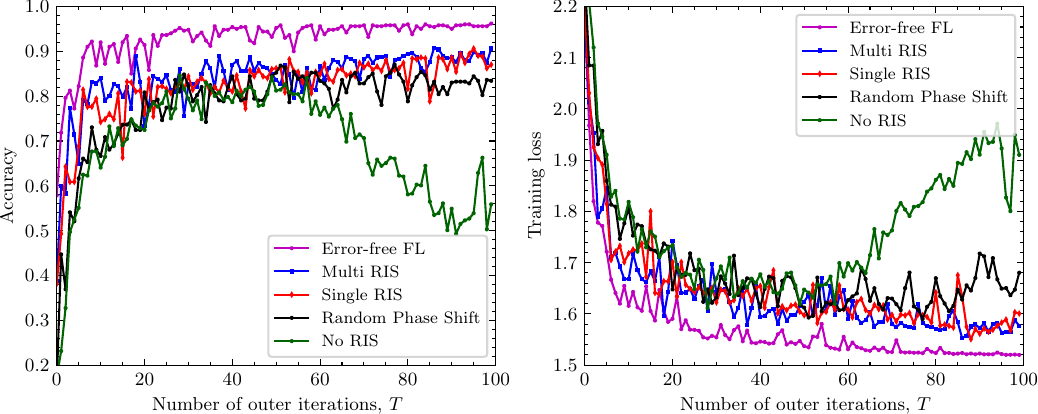} 
			\label{fig:loss}
		}
		\caption{The learning performance over different communication rounds with MNIST dataset.}
		\label{fig:performance}
	\end{figure*}

	\section{Conclusion}

This paper presented a novel framework integrating NOMA and AirFL with multiple RIS to address spectrum scarcity and support diverse communication and learning tasks. Leveraging the reconfigurability of RISs, the framework enhances channel conditions and supports concurrent uplink transmissions from heterogeneous users. We formulated the joint optimization of the denoising factor, power allocation, and RIS configuration as a non-convex problem, addressing key challenges such as co-channel interference and imperfect CSI and SIC. To solve this, we recast the problem as an MDP and employed an LSTM-DDPG algorithm. The proposed method demonstrated superior performance over benchmarks, including DDPG, SAC, and A2C, particularly in addressing channel uncertainties. These findings highlight the potential of the RIS-aided NOMA-AirFL framework as a promising solution for next-generation wireless networks.

	\section*{Appendix}
	%\section*{Proof of Theorem 1}

	Consider a synchronization process carried out for \( t \) rounds, where the difference in loss between consecutive rounds is  represented as \( F(\boldsymbol{w}^{(t+1)}) - F(\boldsymbol{w}^{(t)}) \). If we let \( \boldsymbol{w} = \boldsymbol{w}^{(t+1)} \) and \( \boldsymbol{v} = \boldsymbol{w}^{(t)} \), by employing (19), we get
	\begin{multline}
		F(\boldsymbol{w}^{(t+1)}) - F(\boldsymbol{w}^{(t)})  \leq \\ \nabla F(\boldsymbol{w}^{(t)})^T (\boldsymbol{w}^{(t+1)} - \boldsymbol{w}^{(t)}) + \frac{L}{2} \|\boldsymbol{w}^{(t+1)} - \boldsymbol{w}^{(t)}\|_2^2.
	\end{multline}
Following the model update rule specified in (4), the global model is updated as:
	$\boldsymbol{w}^{(t+1)} = \boldsymbol{w}^{(t)} - \lambda \hat{\boldsymbol{g}}^{(t)},$
	where \( \hat{\boldsymbol{g}}^{(t)} \) signifies the aggregated global gradient defined in (23). Let $\boldsymbol{\epsilon}^{(t)}=\hat{\boldsymbol{g}}^{(t)}-{\boldsymbol{g}}^{(t)}$ denote the aggregation error at round $t$. Consequently, based on (4) and (23), the random aggregation error induced by OTA aggregation is expressed as:
	\begin{multline}
\boldsymbol{\epsilon}^{(t)}=\hat{\boldsymbol{g}}^{(t)}-{\boldsymbol{g}}^{(t)}= {\frac{1}{K\sqrt{\eta}}  \sum_{k=1}^{K} \left( (\hat{h}_k^{(t)}+\varepsilon_{h,k}^{(t)}
			) \sqrt{p_k^{(t)}}-1 \right)\boldsymbol{g}_k^{(t)}} \\+ {\frac{\sqrt{\epsilon_b}}{K\sqrt{\eta}} \sum_{n=K+1}^{K+N} (\hat{h}_n^{(t)}+\varepsilon_{h,n}^{(t)}
			) \sqrt{p_n^{(t)}} s_n^{(t)}} + {\frac{z_0^{(t)}}{K\sqrt{\eta}}}.
	\end{multline}
	By defining $\boldsymbol{g}^{(t)} = \nabla F(\boldsymbol{w}^{(t)})$, and substituting before mentioned equality, we have:
	\begin{multline}
		F(\boldsymbol{w}^{(t+1)}) - F(\boldsymbol{w}^{(t)}) \leq (\boldsymbol{g}^{(t)})^T (-\lambda \hat{\boldsymbol{g}}^{(t)}) + \frac{L}{2} \|- \lambda \hat{\boldsymbol{g}}^{(t)}\|_2^2 \\=(\boldsymbol{g}^{(t)})^T (-\lambda ({\boldsymbol{g}}^{(t)}+\boldsymbol{\epsilon}^{(t)})) + \frac{L}{2} \|- \lambda ({\boldsymbol{g}}^{(t)}+\boldsymbol{\epsilon}^{(t)})\|_2^2 \\=-\lambda (\boldsymbol{g}^{(t)})^T ({\boldsymbol{g}}^{(t)}+\boldsymbol{\epsilon}^{(t)})+ \frac{L\lambda^2}{2} \| {\boldsymbol{g}}^{(t)}+\boldsymbol{\epsilon}^{(t)}\|_2^2.
	\end{multline}
	By applying the expectation on both sides of (41), we obtain
	\begin{multline}
		\mathbb{E}\left( F(\boldsymbol{w}^{(t+1)}) - F(\boldsymbol{w}^{(t)}) \right) \leq -\lambda (\boldsymbol{g}^{(t)})^T \mathbb{E}\left[{\boldsymbol{g}}^{(t)}+\boldsymbol{\epsilon}^{(t)}  \right]\\+ \frac{L\lambda^2}{2} \mathbb{E}\left[\| {\boldsymbol{g}}^{(t)}+\boldsymbol{\epsilon}^{(t)}\|_2^2\right]=-\lambda \| {\boldsymbol{g}}^{(t)}\|_2^2-\lambda (\boldsymbol{g}^{(t)})^T \mathbb{E}\left[\boldsymbol{\epsilon}^{(t)}  \right]\\+\frac{L\lambda^2}{2}\mathbb{E}\left[\| {\boldsymbol{g}}^{(t)}\|_2^2\right]+L\lambda^2(\boldsymbol{g}^{(t)})^T\mathbb{E}\left[\boldsymbol{\epsilon}^{(t)}  \right]+\frac{L\lambda^2}{2}\mathbb{E}\left[\| \boldsymbol{\epsilon}^{(t)}\|_2^2\right]\\\le -\lambda \| {\boldsymbol{g}}^{(t)}\|_2^2-\lambda (\boldsymbol{g}^{(t)})^T \mathbb{E}\left[\boldsymbol{\epsilon}^{(t)}  \right]+\frac{L\lambda^2}{2}\left( \| {\boldsymbol{g}}^{(t)}\|_2^2+\| \boldsymbol{\delta}\|_2^2 \right)\\+L\lambda^2(\boldsymbol{g}^{(t)})^T\mathbb{E}\left[\boldsymbol{\epsilon}^{(t)}  \right]+\frac{L\lambda^2}{2}\mathbb{E}\left[\| \boldsymbol{\epsilon}^{(t)}\|_2^2\right],
	\end{multline}
	where the last equality is due to Assumption 3. By applying the inequality of arithmetic and geometric means and the Cauchy Schwarz inequality, i.e., $\mp x_{1}^Tx_{2}\le \| x_{1}\|\| x_{2}\|\le \frac{1}{2}\| x_{1}\|^2+\frac{1}{2}\| x_{2}\|^2$, we further have
	\begin{multline}
		\mathbb{E}\left( F(\boldsymbol{w}^{(t+1)}) - F(\boldsymbol{w}^{(t)}) \right) \leq  
		-\lambda \| {\boldsymbol{g}}^{(t)}\|_2^2+\frac{\lambda^2 }{2}\| {\boldsymbol{g}}^{(t)}\|_2^2\\+\frac{1}{2}\| \mathbb{E}[\boldsymbol{\epsilon}^{(t)}]  \|_2^2+\frac{L\lambda^2}{2}\left( \| {\boldsymbol{g}}^{(t)}\|_2^2+\| \delta\|_2^2 \right)+\frac{L^2\lambda^2}{2}\| \mathbb{E}[\boldsymbol{\epsilon}^{(t)}]  \|_2^2\\+\frac{\lambda^2 }{2}\| {\boldsymbol{g}}^{(t)}\|_2^2+\frac{L\lambda^2}{2}\mathbb{E}\left[\| \boldsymbol{\epsilon}^{(t)}\|_2^2\right]
		=-\left( \lambda-\lambda^2-\frac{L\lambda^2}{2} \right)\| {\boldsymbol{g}}^{(t)}\|_2^2\\+\frac{L\lambda^2}{2}\| \boldsymbol{\delta}\|_2^2 +\frac{1+L^2\lambda^2}{2}\| \mathbb{E}[\boldsymbol{\epsilon}^{(t)}] \|_2^2+\frac{L\lambda^2}{2}\mathbb{E}\left[\| \boldsymbol{\epsilon}^{(t)}\|_2^2\right]
		\\\le -\frac{\lambda}{2}\| {\boldsymbol{g}}^{(t)}\|_2^2+\frac{L\lambda^2}{2}\| \delta\|_2^2 +\frac{1+L^2\lambda^2}{2}\| \mathbb{E}[\boldsymbol{\epsilon}^{(t)}] \|_2^2\\+\frac{L\lambda^2}{2}\mathbb{E}\left[\| \boldsymbol{\epsilon}^{(t)}\|_2^2\right],
	\end{multline}
	where the last inequality follows that $\lambda\le \frac{2}{2+L}$. Subtracting $F(w^{*})$ from both sides of (18) yields an upper bound for the expected performance gap in the $t$-th communication round:
	\begin{equation}
		\mathbb{E}\left( F(\boldsymbol{w}^{(t+1)})\right) - F(\boldsymbol{w}^{*})\le (1-\lambda\mu)\left(  F(\boldsymbol{w}^{(t)}) - F(\boldsymbol{w}^{*}) \right)+\psi^{(t)},
	\end{equation}
	where, $\psi^{(t)}=\frac{L\lambda^2}{2}\| \boldsymbol{\delta}\|_2^2 +\frac{ 1+L^2\lambda^2 }{2}\| \mathbb{E}[\boldsymbol{\epsilon}^{(t)}] \|_2^2+\frac{L\lambda^2}{2}\mathbb{E}\left[\| \boldsymbol{\epsilon}^{(t)}\|_2^2\right]$. Recursively applying this bound over $T$ rounds gives the cumulative optimality gap:
	\begin{multline}
		\mathbb{E}\left[ F(\boldsymbol{w}^{(T+1)}) - F(\boldsymbol{w}^*) \right]  \leq (1-\lambda\mu)\left(  F(\boldsymbol{w}^{(T)}) - F(\boldsymbol{w}^{*}) \right) \\+ \psi^{(T)} 
		\leq (1-\lambda\mu) \left( (1-\lambda\mu) \left( F(\boldsymbol{w})^{(T-1)} - F(\boldsymbol{w}^*) \right) + \psi^{(T-1)} \right)\\+ \psi^{(T)}  
		\leq \ldots 
		\leq (1-\lambda\mu)^{(T)}(F(\boldsymbol{w}^{(1)})-F(\boldsymbol{w}^{*}))\\+\sum_{t=1}^{T}(1-\lambda\mu)^{(T-t)}\psi^{(t)}. 
	\end{multline}
	This complete the proof.

	\bibliographystyle{IEEEtran}
	\bibliography{OTA_FL}

	\vfill
	
\end{document}